\documentclass[12pt]{article}

\usepackage{scicite}
\usepackage{times}
\usepackage{graphicx}
\usepackage[hidelinks]{hyperref}
\usepackage{xcolor}
\usepackage{pdfpages}
\usepackage{longtable}
\usepackage{gensymb}
\usepackage{subcaption}
\usepackage{booktabs} 
\usepackage{amssymb}
\usepackage{verbatim}

\newenvironment{sciabstract}{%
\begin{quote} \bf}
{\end{quote}}

\newcommand{\vecP}{\mathbf{p}}

\title{Assessing the Risk of Permafrost Degradation with Physics-Informed Machine Learning} 

\author{
Polina Pilyugina, Timofey Chernikov, Alexey Zaytsev, \\Alexander Bulkin, Evgeny Burnaev, Ilya Belalov, \\Nazar Sotiriadi, Yury Maximov, Oleg Anisimov\\ [1ex]
}

\date{\today}

\begin{document} 


\baselineskip24pt


\maketitle

\begin{sciabstract}
  Global warming accelerates permafrost degradation, impacting the reliability of critical infrastructure used by more than five million people daily. Furthermore, permafrost thaw produces substantial methane emissions, further accelerating global warming and climate change and putting more than eight billion people at additional risk. 
  To mitigate the upcoming risk, policymakers and stakeholders must be given an accurate prediction of the thaw development. Unfortunately, comprehensive physics-based permafrost models require location-specific fine-tuning that is challenging in practice. Models of intermediate complexity require few input parameters but have relatively low accuracy. The performance of pure data-driven models is low as well as the observational data is sparse and limited.
  In this work, we designed a physics-informed machine-learning approach for permafrost thaw prediction. The method uses a heat equation to regularize 
  data-driven approach trained over permafrost monitoring data and climate projections. The latter leads to higher precision and better numerical stability allowing for reliable decision-making for construction and maintenance in the areas endangered by permafrost thaw with a time horizon of decades.
\end{sciabstract}

\section*{Introduction}\label{subsec:intro}
A distinctive feature of the Arctic environment is perennially frozen ground, or permafrost, which is defined as any subsurface material that remains below $0^{\circ}$C for more than two consecutive years.  
Permafrost can be found on land, in high mountains and even on the shelf of some Arctic seas. Collectively, these three types of permafrost currently occupy about a quarter of the Northern Hemisphere land surface, including about 16.7 million km$^2$ in Eurasia, 10.2 million km$^2$ in North America~\cite{Obu2021}; see. Fig.~\ref{fig:map_permafrost_CALM} for details.

The Arctic Circumpolar Permafrost Region encompasses 1162 permafrost settlements~\cite{Ramage2021}, accommodating approximately 5 million inhabitants.
Many objects of critical infrastructure are located on permafrost \cite{Hjort2022}, including freeways, railroads, oil and gas pipelines, nuclear stations, 
which have been under the effect of permafrost thaw \cite{dorecost,larsen2008estimating,melvin2017climate,ramage2021population}. 
For example, according to the Circum-Arctic Map of Permafrost and Ground Ice Conditions by the International Permafrost Association (IPA Map) \cite{Ferrians2002}, permafrost comprises almost 65\% of Russian territory, 80\% of Alaska \cite{map_alaska} and about 50\% of Canada \cite{ferrians1994permafrost}.
Permafrost severely affects many aspects of daily life and industrial processes, \textit{e.g.}, mining, in these regions~\cite {Streletskiy2015a,Streletskiy2015b,Streletskiy2021} and provide a substantial risk to the global supply chain~\cite{alexander2022managing}. 
Furthermore, most of the critical infrastructure and houses in high-latitudes are built on permafrost~\cite{Streletskiy2019}, which makes it crucial to control and monitor the thaw.

The permafrost is characterized by the two major parameters: active layer thickness and ground temperature. The active layer is the uppermost soil above the permafrost, which thaws in the summer and re-freezes in the winter. 
The active layer is the soil above the permafrost, which thaws in the summer and freezes in the winter. The active layer thickness (ALT) is defined as the maximal seasonal thaw depth and measured at the end of a summer season. The mean annual ground temperature (MAGT) is defined as the temperature at the depth where the annual temperature amplitude is less than 0.1\degree C.
Conventional permafrost degradation models connect the climate parameters, particularly air temperature and precipitation, with the ALT and MAGT. Soil and vegetation properties largely govern this relationship, and most permafrost models explicitly include them. Both ALT and MAGT are vulnerable to climate change. Global warming results in a higher ground temperature, which causes permafrost thaw and increases the active layer thickness. 
These changes in the permafrost state negatively affect the stability of buildings and infrastructures and decrease the bearing capacity of the building foundations~\cite{Anisimov2010,Anisimov2016_georisks,Streletskiy2015a,Streletskiy2015b}. 

There is growing observational evidence that such effects are already taking place. 
Various studies~\cite{Streletskiy2015a,Streletskiy2015b,Streletskiy2021,Anisimov2010} report the effect of global warming on the destruction of buildings and infrastructure built on permafrost. 
Permafrost thaw also forms gas emission craters, which release large amounts of methane and other gases into the atmosphere~\cite{bogoyavlensky2014threat,khimenkov2019structural}.
Permafrost carbon feedback, a phenomenon of significant concern, involves the release of substantial soil carbon deposits through permafrost thaw, amplifying the impacts on ecosystems ~\cite{natali2019large, schuur2015climate, Mattsson2007, Christensen2004, Miner2022}. Additionally, it is worth noting that methane emissions, with a considerably stronger greenhouse effect compared to an equal amount of CO$_2$, pose particular concerns in this context.
Other studies in this field outline economic consequences of climate change related to permafrost. 
In \cite{Porfiriev2017}, the authors approach this problem from the point of its influence on the macroeconomic variables. 
Studies \cite{LARSEN2008442, Melvin2017} have estimated that the cost of critical infrastructure damage due to permafrost thaw in North America could reach about \$5 billion by the end of the 21$^{st}$ century.
A similar study for Northern Europe and Asia is given in \cite{Streletskiy2021, Hjort2018, Anisimov2016_georisks, Shiklomanov2017, Badina2021}. 

Mitigating the consequences of permafrost degradation requires a method to estimate the permafrost characteristics under various climate change scenarios.
This paper focuses on predicting permafrost changes related to climate change in permafrost-rich areas and combines physics-based mathematical models with data-driven methods aiming at improving their quality over limited data.

\section*{Contribution}

This paper presents a novel physics-informed machine-learning approach for permafrost thaw prediction to address the challenges posed by global warming and its impact on critical infrastructure and climate change. The key contributions of this research are as follows:
\begin{itemize}
    \item \emph{a novel robust and accurate prediction framework} combining the strengths of physics-based permafrost degradation models and machine learning. The solution is properly regularized and constrained by fundamental physical principles;
    \item \emph{state-of-the-art prediction accuracy} substantially improving existing algorithms and supporting risk assessment and decision-making;
    \item \emph{extensive empirical validation and uncertainty estimation} over  Circumpolar Active Layer Monitoring (CALM) observational data. 
\end{itemize}


Overall, our physics-informed machine-learning approach offers a powerful tool for predicting permafrost thaw, mitigating risks to critical infrastructure, and understanding the implications of permafrost degradation on global warming and climate change. This research contributes to the growing field of climate science and offers valuable insights for infrastructure development in high-latitude regions facing the challenges of permafrost thaw.

\section*{
Algorithms and Models}

A compendium of methodologies in permafrost modeling is available in~\cite{Riseborough2008}. A simplest semi-empirical approach considers two key parameters, ALT and MAGT, which are linked statistically to the climatic parameters using historical observational data. Permafrost observations are spatially and temporally sparse and irregular, which poses additional challenges to a researcher. The lack of a unified standard for the observations further complicates the problem.

Another type of permafrost models \cite{Anisimov2016_model} takes into account the climatic parameters and variability of snow, vegetation, and soil properties. The results demonstrate a good fit with the observational data.
Such a model is used in~\cite{Anisimov2016_georisks} to estimate the geocryological risks of permafrost thawing.
A similar model was used to simulate the temperature of the permafrost and depth of a seasonally thawed layer in \cite{Streletskiy2019}. The authors used kriging to distribute the values over a spatial grid with a resolution of $0.25^{\circ}$. Predictive calculations were made with the scenario of climate change derived from six CMIP 5 (Coupled Model Intercomparison Project Phase) Earth System Models under the assumption of the continuing growth of greenhouse gas emissions (so-called Representative Concentration Pathway (RCP) 8.5 scenario). Streletskiy \textit{et al.} in~\cite{Streletskiy2019} compared two periods: the present (2005-2015) and the middle of the 21st century~(2050-2059).

A different \textit{numerical} approach based on statistical models was described in \cite{Hjort2018}. Hjort \textit{et al.} computed characteristics such as the temperature, precipitation, organic carbon content in the soil, soil type, water bodies, solar radiation, and topographic properties for each point of the grid. These factors were predicted with the models from CMIP 5 climate scenario. The variables depend on greenhouse gas concentrations. Three scenarios adopted by the IPCC (RCP 2.6, RCP 4.5 and RCP 8.5) resulted in different values for the variables. In further analysis, these factors served as predictors for each scenario of greenhouse gas concentrations. The analysis included four statistical models (generalized linear model (GLM), generalized additive model (GAM), random forest (RF), and generalized boosting model (GBM)) and their ensemble. The analysis resulted in a map of soil the temperature and thickness of the seasonally thawed layer.
The prediction uncertainty of the model based on the ensemble of four methods was estimated as $\pm0.77^{\circ}$C for soil temperature and $\pm$ 37 cm for the thickness of the seasonally thawed layer. 

An analogous approach was considered in \cite{aalto2018} with more focus on the modeling procedure. The predictive error (root mean squared error) of this approach, which was evaluated on the hindcast data for past periods, was reported to be 53 cm for ALT and 1.58\degree C for MAGT. To date, to the best of our knowledge, this is the best precision achieved in permafrost thaw prediction. However, a model with a prediction error of 53 cm is a poor guide for maintaining a structure built over permafrost.

Ultimately, the result of \cite{shirley2023machine} imposes the inability of machine learning methods to predict environmental processes, including the permafrost thaw, at high latitude carbon cycle balances. The crucial contribution of this paper is regularizing the classical machine learning approaches with a special version of the heat equation \cite{shirley2023machine}. The latter substantially improves the prediction quality of permafrost thaw and opens the door to efficient yet accurate environmental risk assessment models in high latitudes.  

There are many approaches for permafrost modeling and prediction, which may or may not include the effects of climate change, but the quality of the models limits their use in applications related to the maintenance of buildings and construction design for the long-term sustainable development of the northern regions.

In this study, we focus on spatial models and their interaction with 
climatic projections for the future. 
Kudryavtsev equilibrium model \cite{kudryavtsev1977fundamentals, shiklomanov1999analytic, Anisimov2016_model} is particularly useful. Equilibrium models suggest that the MAGT is in balance with the atmospheric parameters. They have relatively low data requirements; they use the mean monthly temperature and precipitation data as climate forcing with a few edaphic parameters that characterize the soil thermal properties, snow, and vegetation. The equilibrium model developed by V. Kudryavtsev is one of the most successful examples \cite{kudryavtsev1977fundamentals}. With a slight modification, the model has been used in many subsequent studies \cite{jafarov2012, sazonova2003model, Anisimov2016_georisks, Anisimov2016_model, Streletskiy2012, Streletskiy2019}. 

Despite the recent improvement in the availability of observational MAGT and ALT data through dedicated web-portal \cite{gtnpalt}, permafrost modeling remains data-limited. Thus, more powerful transient models cannot be effectively used because all necessary forcing data and edaphic parameters are often not available, and the model cannot be appropriately calibrated. Given these limitations, in this study we used the equilibrium permafrost model.

The model is based on the numerical solution of the nonlinear parabolic equation by Kudryavtsev \cite{kudryavtsev1977fundamentals}. Kudryavtsev model accounts for the snow, vegetation, organic and mineral substrate, and variable thermal properties \cite{shiklomanov1999analytic,sazonova2003model}. The model performs well in extensive validations by empirical observations \cite{shiklomanov1999analytic}.

In the Kudryavtsev model, every grid element (a pixel on the map) is associated with observable variables such as the snow cover, vegetation, and soil characteristics. 
The model outputs the ground temperature and active layer thickness. In this study, we employed a modification of the Kudryavtsev model in \cite{Anisimov2016_model}. For a detailed description of the model, see \cite{riseborough2008recent, Anisimov2016_model}.

The input parameters of the model include characteristics of the atmosphere (monthly air temperatures and monthly precipitation) for climatic forcing. Additionally, the model configuration includes various soil specifications. By varying the main parameter --- the type of soil (\textit{i.e.}, clay, loan, peat) --- we obtained an ensemble of models. Other parameters, \textit{i.e.},, the snow cover and organic layer depth, were held constant. The organic layer depth is defined based on the persistent vegetation type in the area.

\section*{Materials and Methods}
The key idea of the physics-informed machine learning approach for permafrost thaw degradation 
is to boost the prediction quality by combining state-of-the-art data-driven methods with physical-based models. The latter allows for building a high-fidelity thaw predictor over limited observations data. 

\begin{figure}[h]
    \centering
    \includegraphics[scale=0.5]{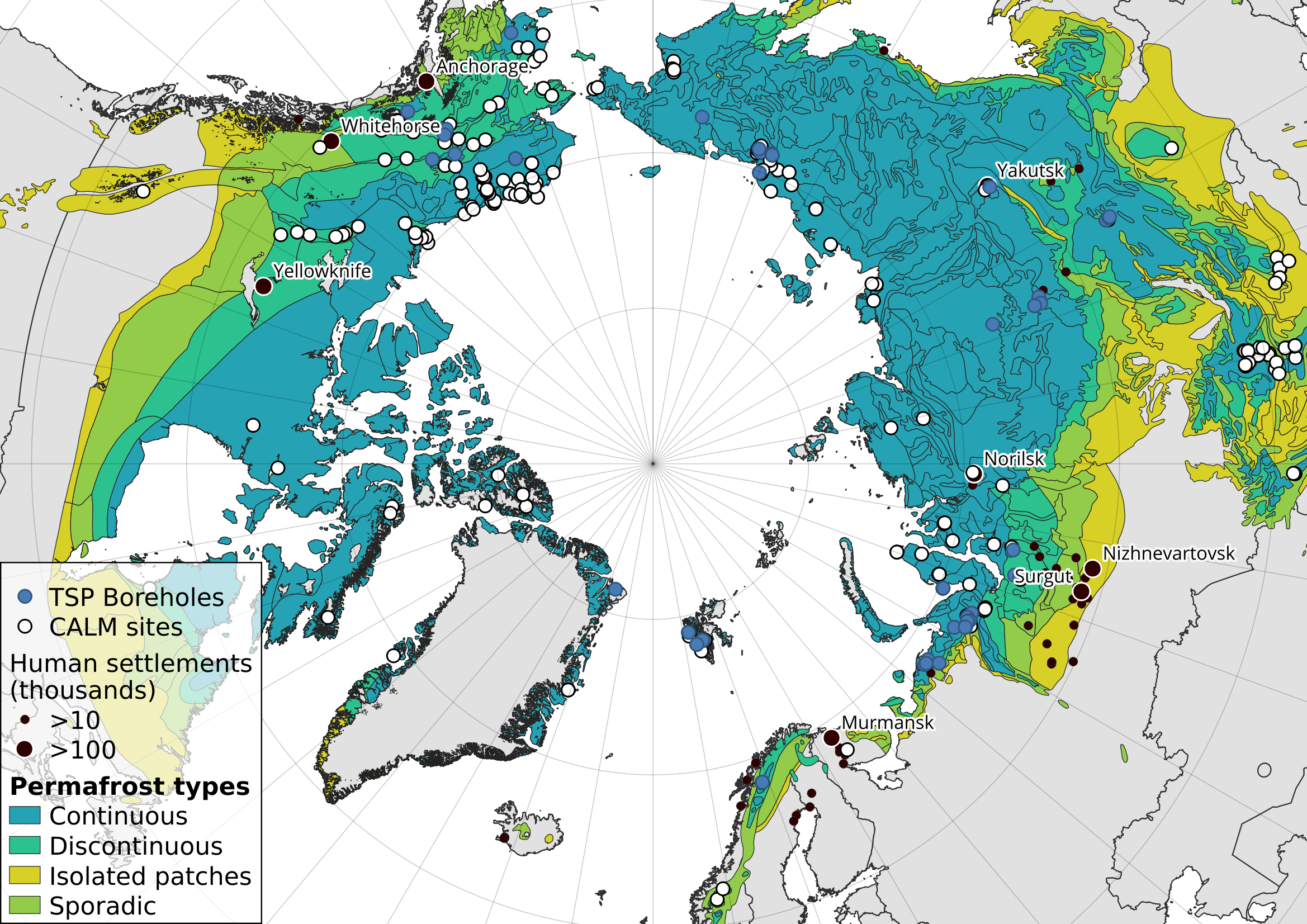}
    \caption{Distribution of permafrost types according to the IPA Permafrost Map with CALM sites and TSP boreholes.}
    \label{fig:map_permafrost_CALM}
\end{figure}

The Circumpolar Active Layer Monitoring (CALM) program includes data from 265 sites in the Northern Hemisphere in 15 countries \cite{CALM}. Figure~\ref{fig:map_permafrost_CALM} demonstrates the locations of the CALM sites.
As a part of the approach to assessing the risks of permafrost degradation, the seasonal thaw depth is modeled for all points of the grid. 

\begin{figure}[h]
    \centering
    \includegraphics[width=16cm]{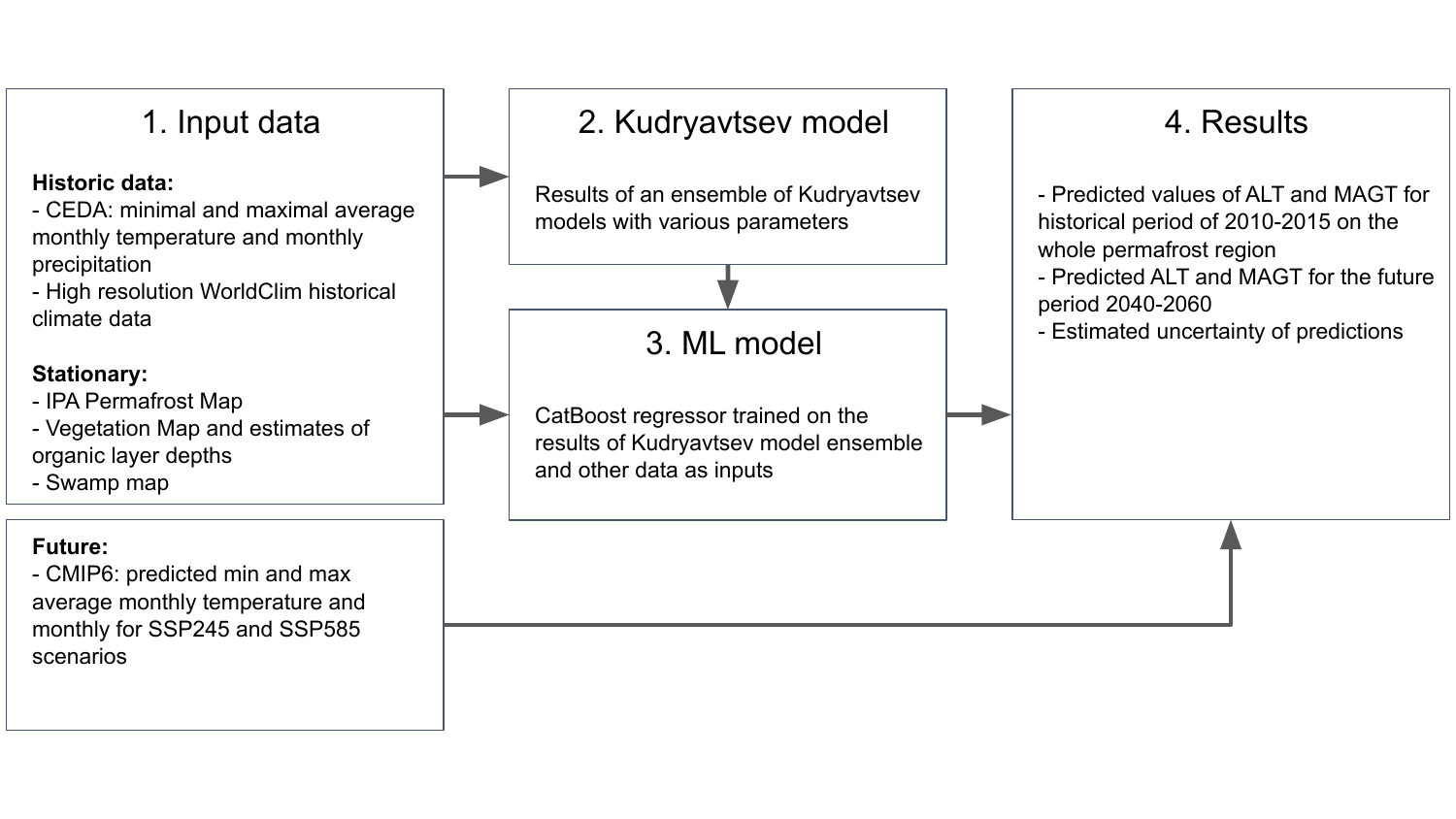}
    \caption{Model flowchart. Graphical summary of the \textbf{methods} section. }
    \label{fig:perm_model_schema}
\end{figure}

We trained a machine learning model based on gradient boosting using the data from the CALM sites. 
The input data included dynamically measured parameters for climate, static data on soil and vegetation, and results for an ensemble of the Kudryavtsev model for different initializations of the parameters. 
Figure \ref{fig:perm_model_schema} shows the model flowchart.
The first step consists of processing the input data and submitting the data to Kudryavtsev model \cite{kudryavtsev1977fundamentals}.

Kudryavtsev model is a solution to the equation of heat flow theory \cite{Riseborough2008}, described by Eq.~\ref{eqn:heatflow}.
\begin{equation}\label{eqn:heatflow}
    C \frac{\delta T}{\delta t} = k \frac{\delta^2t}{\delta z^2}
\end{equation}
where $C$ is volumetric head capacity in Jm$^{-3}$, $T$ is temperature, $t$ is time and $z$ is depth in meters.
The mathematical setup of the model is the following.
The depth of seasonal thawing, $Z_{thaw}$, is calculated using the semi-empirical equation: 
\begin{equation}
Z_{thaw} = \frac{2(A_{s}-T_{z})\cdot \left[\frac{\lambda \cdot P_{sn} \cdot C}{\pi}\right]^{1/2}+\frac{(2A_{z} \cdot C \cdot Z_{c} + Q_{ph} \cdot Z_{c})\cdot Q_{ph} \cdot \left[\frac{\lambda \cdot P_{sn}}{\pi \cdot C} \right ]^{1/2}}{2A_{z} \cdot C \cdot Z_{c} + Q_{ph} \cdot Z_{thaw} + (2A_{z}\cdot C + Q_{ph}) \cdot \left[\frac{\lambda \cdot P_{sn}}{\pi \cdot C} \right ]^{1/2}}}{2A_{z}\cdot C + Q_{ph}}
\end{equation}
where
\begin{equation}
    A_{z} = \frac{A_{s}-T_{z}}{\ln \left[\frac{A_{s}+Q_{ph} / 2C}{T_{z}+Q_{ph}/2C}\right]} - \frac{Q_{ph}}{2C}
    \qquad \textrm{and}\qquad 
    Z_{c} = \frac{2(A_{s}-T_{z}) \cdot \sqrt{\frac{\lambda \cdot P_{sn} \cdot C}{\pi}}}{2A_{z}\cdot C + Q_ph}.
\end{equation}
Here, $A_{s}$ is the annual amplitude of the soil-surface temperature (in  ${}^{\circ}$ C), $T_{z}$ is the mean annual temperature at the depth of seasonal thawing (in ${}^{\circ}$ C), $\lambda$ is the thermal conductivity of soil in the thawed state ($Wm^{-1}K^{-1}$), $C$ is the volumetric heat capacity of soil in the thawed state ($Jm^{-3} K^{-1}$), $P_{sn}$ is the period of the temperature wave (sec), and $Q_{ph}$ is the volumetric latent heat of phase changes ($Jm^{-3}$).

Then, we ran Kudryavtsev model with various initializations of parameters to obtain the ALT and MAGT values for each initialization. 
These values were forwarded as inputs to a Machine Learning (ML) model with other data, as described in the \textbf{Data} section.
The ML model was subsequently trained and evaluated using available historical observations from the CALM and TSP stations.
We used gridded climate data from the ensemble of CMIP6 Earth System Models to obtain predictions for the periods of 2010 -- 2015 and 2040 -- 2060. We considered two greenhouse-gas emission scenarios defined by so-called Shared socio-economical pathways (SSP). SSP-scenarios consider the impact of climate change on the development of societies and economies~\cite{riahi2017shared}.


\subsection*{Data}\label{sec:data}

\begin{table}[ht]
    \centering
    \begin{tabular}
    {|l|l|l|l|}
        \hline
        {\bf Input data} & {\bf Parameters} &  {\bf Units} & {\bf Period} \\
        \hline
        IPA Permafrost Map & Permafrost type & Categorical & N/A \\
        \hline
     swamp & Proportion of swamps in the area & \% & N/A \\
        \hline
      CEDA & Temperatures, precipitation & $^{\circ}$C, mm & 1901-2020 \\
        \hline
    WorldClim (historical) & Temperature, precipitation &  $^{\circ}$C, mm & 1960-2018 \\
        \hline
  CALM & Thaw depth & cm & 1969-2021 \\
  &&& (irregular data) \\
        \hline
     WorldClim (CMIP6) & Temperature, precipitation & $^{\circ}$C, mm & 2006-2100 \\
        \hline
vegetation & Type of vegetation & Categorical & N/A \\
        \hline
    GTNP-TSP & Temperature, zero-amplitude depth &$^{\circ}$C, cm & 1901-2020 \\
        \hline
    \end{tabular}

    \caption{Input data}
    \label{tab:datasets}
\end{table}

The spatial resolution for historical data was 0.5 angle degrees, except for the GTNP-ALT dataset (see the description below). 
We used high-resolution temperature and precipitation data from WorldClim\cite{fick2017worldclim,harris2014updated} for future predictions and historical assessment. The temporal resolution for the time series data varied from 1 month to 1 year. We aggregated values of interest for each pixel with the datasets detailed below:
\begin{description}
    \item[CEDA] \cite{ceda}\\
    Various climate variables: the cloud cover (average fraction of the sky obscured by clouds), daily temperature range, proportion of time with negative temperature in a 24-hour period, precipitation, monthly average daily minimum temperature (the minimum temperature was taken daily and averaged over the month), monthly average maximum daily temperature, monthly average temperature, proportion of time during which precipitation occurs. The temporal resolution for the data is 1 month.
    
    \item[CALM] \cite{gtnpalt}\\
    Statistics of the thaw depth collected at CALM sites (Figure \ref{fig:map_permafrost_CALM}). Each data point has eight entries. Seven of those describe the characteristics of the thaw depth: average, median, minimum, maximum, 25th percentile, 75th percentile, and standard deviation. For each point, measurements were performed at the vertices of a rectangular grid of 1 km $\times$ 1 km with 100-m steps; there were 121 measurements in total. In some cases, several measurements were missing. Data from some points were irregularly collected or during short time intervals. To bring the data to a common format with a spatial resolution of 0.5 angle degrees, we considered the nearest CALM site for each point on the map. This distance is the eighth entry for each data point. The maximum distance between the CALM site and the corresponding point on the map was 200 km (4 cases). The temporal resolution for this dataset is 1 year.
    
    \item[WorldClim (historical)] \cite{fick2017worldclim}\\
    High-resolution historical climatic data projected from CRU-TS-4.03 using WorldClim 2.1 for bias correction (Climatic Research Unit, University of East Anglia). The data cover the period from 1960 to 2018. The spatial resolution of the original dataset is 2.5 minutes ($\sim$21 km$^2$). We projected it to the 0.5-angular-degree grid. The time resolution is 1 month.
    
    \item[IPA] \cite{IPAMap}\\
    Map of the distribution of permafrost types. We rasterized the original map in the vector format to bring it to a resolution of 0.5 angle degrees to fit the common format for other data (Figure \ref{fig:map_permafrost_CALM}). These data are constant in time.
    
    \item[WorldClim (CMIP6)] \cite{CMIP6data} \\
    The dataset from the CanESM5 model contains data on the air temperature (mean, minimum, and maximum) in $^{\circ}$C and precipitation (mm). The time resolution is 1 month.
    
    \item[swamp]\ \\
    These data were kindly provided by Professor Oleg Anisimov. The data show the percentage of wetlands for each point on the map (see Supplementary Figure \ref{fig:swamp}). The data are constant in time.

    \item[vegetation]\ \\
    These data were kindly provided by Professor Oleg Anisimov. The data describe the types of landscape or biotopes (see Fig. S\ref{fig:veg}). These data are constant in time.

    \item[GTNP-TSP]\ \\
    Zero Annual Amplitude -- the depth at which the temperature change throughout the year was less than $0.1^\circ$ C \cite{romanovsky2010permafrost}. These data describe a thermal state of the permafrost, which was measured at boreholes (TSP boreholes in Figure \ref{fig:map_permafrost_CALM}). Since the exact depth of the ZAA is not explicitly reported in the dataset, we manually determine the ZAA depth for each weather station and for each year using the GTNP dataset \cite{gtnpalt}. Stations with insufficient data to define the ZAA were omitted. Then, we projected the resulting dataset onto a regular two-dimensional grid to create a map. The final dataset includes both temperature at the ZAA level and ZAA depth value.
\end{description}


\subsection*{Data for modeling}
Let $\mathcal{P}$ be the domain. 
This is the part of the Earth's surface underlaid by permafrost. 
We project this two-dimensional locally continuous surface to the grid $\mathbf{P}$ parameterized by two angles $i$ and $j$. The steps for both angular parameters are 0.5 degrees. For each pixel on the map with coordinates $i$ and $j$, we have observations of different features $\vecP$ in different moments of time~$\tau$. The time step is 1 month. Thus, $\mathcal{P}\supset\mathbf{P}=\{\vecP_{ij}^{\tau}\}$, where each observation has spatial and time indices.

Input data $\vecP$ for our model include the dynamic data on the temperature and precipitation and static information on the swampiness, biotopes, and permafrost type:
\begin{equation}
    \vecP_{ij}^{\tau} = \{i, j, \tau, X_{ij}^\tau, X_{ij}, Y_{ij}^\tau\}
\end{equation}
where
\begin{itemize}
    \item[] $i$, $j$ - latitude and longitude of the center of a pixel, for which data are observed
    \item[] $\tau$ - year of the observation
    \item[] $X_{ij}^\tau$ - 12 monthly values of climatic variables (min and max temperatures, monthly precipitation)
    \item[] $X_{ij}$ - stationary parameters, which do not depend on time: swampiness, biotopes, permafrost type
    \item[] $Y_{ij}^\tau = \{alt_{ ij}^\tau, magt_{ij}^\tau\}$ - observed ALT and MAGT values
\end{itemize}

Our model takes $X_{ij}$ and $X_{ij}^\tau$ as the inputs.
The key variables for this study, $alt_{ ij}^\tau, magt_{ij}^\tau$, were only recorded for a subset of the entire grid $\mathbf{P}$. Furthermore, we used this subset for training and validation.

\subsection*{Implementation}
Our model considers both observed meteorological data and output of Kudryavtsev model (see Fig. \ref{fig:perm_model_schema}). The input data consist of 12 monthly measurements for temperature and precipitation $X_{ij}^\tau$, the thickness of an organic layer, and soil type $X_{ij}$.

We used variations of the latter to parametrize Kudryavtsev model. The initialization of Kudryavtsev model contains parameters for four soil types: sand, loam, clay, and peat.
The dominant soil type in the permafrost areas is clay soil, which has similar physicochemical parameters to loam. Therefore, we obtained the model predictions for both soil types. Moreover, we obtained the model predictions for dry and wet soils. For swamp-associated locations, we used the average for the initializations with the regular soil type and peat, which were weighted according to the \textbf{swamp} dataset (see Supplementary Figure). Thus, for each pixel, we obtained four pairs of ALT and MAGT values to initialize Kudryavtsev model.

Our model predicts the values for $alt_{ij}^\tau$ and $ magt_{ij}^\tau$ using $\vecP_{ij}^{\tau} \in \mathbf{P}_{alt}$ and initializations of Kudryavtsev model from the previous step. 

We conducted experiments with various machine Learning models to evaluate their performance and select the most suitable model.
These models are: Linear regression \cite{galton1886regression}, Multi-layer Perceptron regression \cite{hinton1990connectionist}, Random Forest regression \cite{breiman2001random}, Elastic Net regression \cite{zou2005regularization} and CatBoost Regressor \cite{prokhorenkova2018catboost}.
We selected CatBoost Regressor \cite{prokhorenkova2018catboost} as the most reliable method. This model is a supervised learning meta-algorithm that constructs an ensemble of decision trees.
The CatBoost Regressor reduces the bias and variance in predictions to produce a high-quality and diverse ensemble. 

\begin{figure}
     \centering
     \begin{subfigure}[b]{7.5cm}
         \centering
         \includegraphics[width=0.97\textwidth]{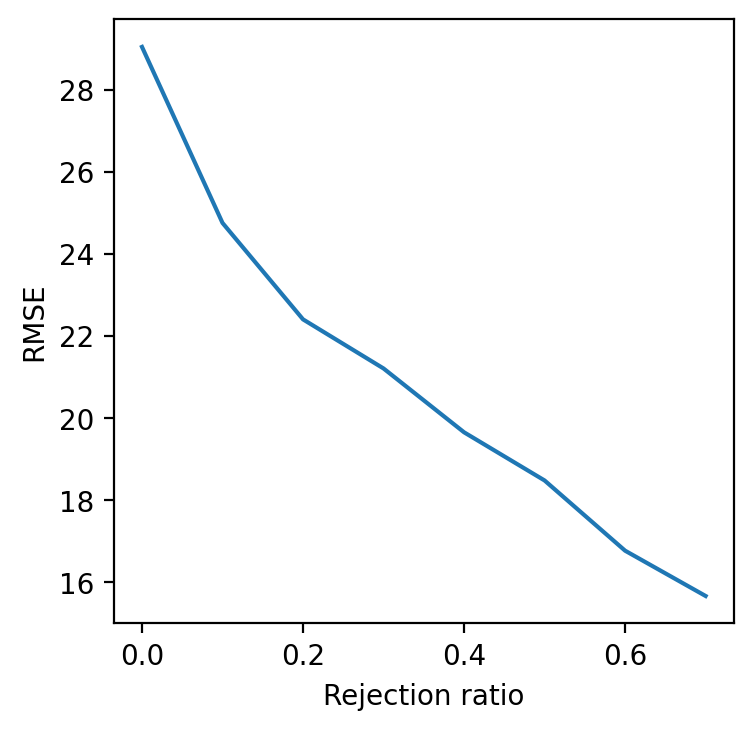}
         \caption{Rejection curve for the ALT model}
     \end{subfigure}
     \hfill
     \begin{subfigure}[b]{7.5cm}
         \centering
         \includegraphics[width=\textwidth]{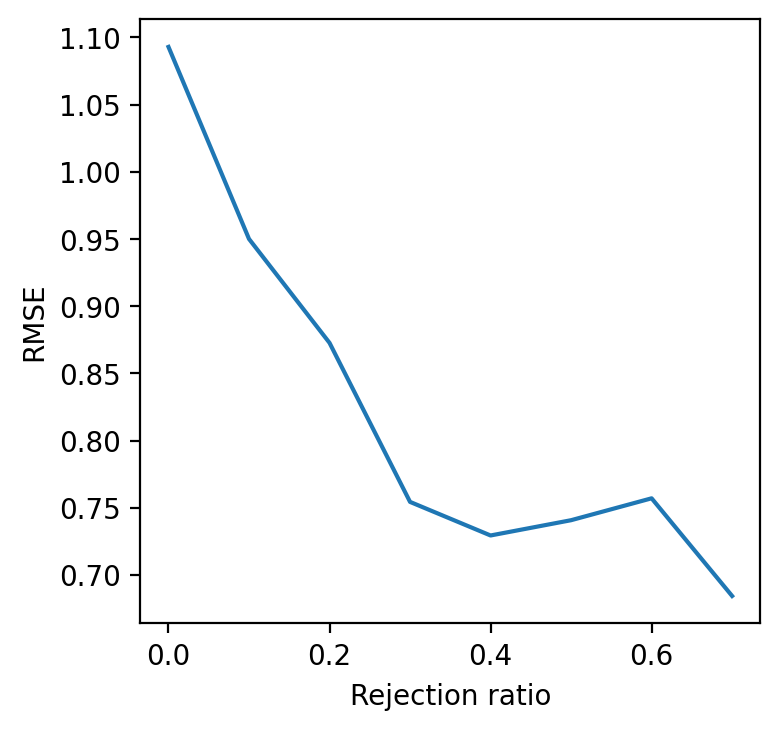}
         \caption{Rejection curve for the MAGT model}
     \end{subfigure}
     \hfill
        \caption{Dependency of the error of the model on the test set alterations with dropping top $k$ \% (rejection ratio) values with the highest predictive uncertainty.}
        \label{fig:uncertainty}
\end{figure}

For uncertainty estimation, we used the algorithms implemented in Catboost. We trained the model using a specific loss function, which enables uncertainty estimation \textit{via} virtual ensembles, according to the original study \cite{malinin2020uncertainty}. We also provided rejection curves for the estimated uncertainty values based on datasets that were depleted of points with the highest predictive uncertainty (Fig. \ref{fig:uncertainty}).

Predictions of permafrost degradation in the future, which is the purpose of our model development, uses data on the temperature and precipitation from the ensemble of CMIP6 earth system models instead of historical data. Therefore, results for Kudryavtsev model were updated according to CMIP6 data.

\section*{Results}

\begin{figure}
     \centering
     \begin{subfigure}[b]{7.5cm}
         \centering
         \includegraphics[width=\textwidth]{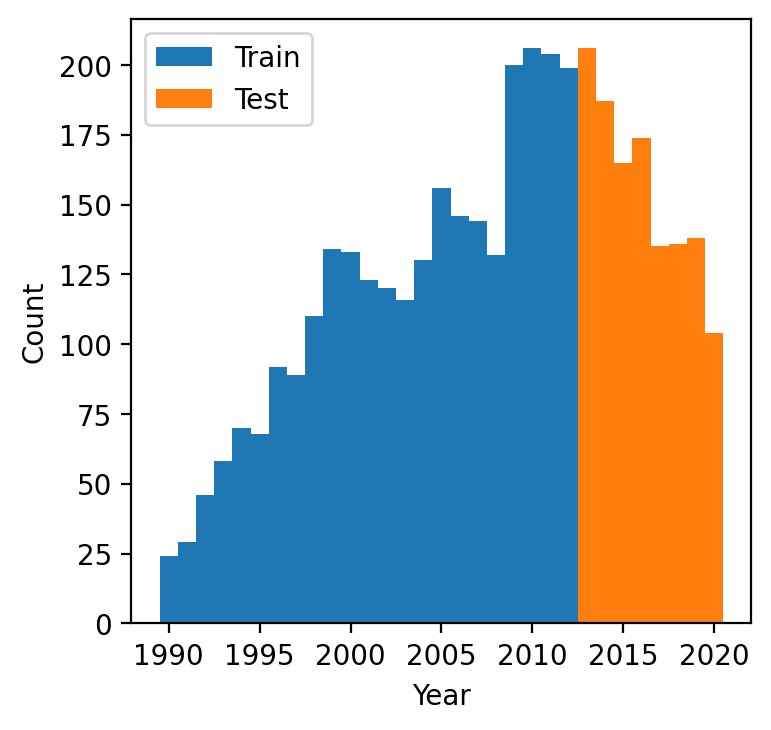}
         \caption{Training-testing split and number of observed ALT values}
     \end{subfigure}
     \hfill
     \begin{subfigure}[b]{7.5cm}
         \centering
         \includegraphics[width=\textwidth]{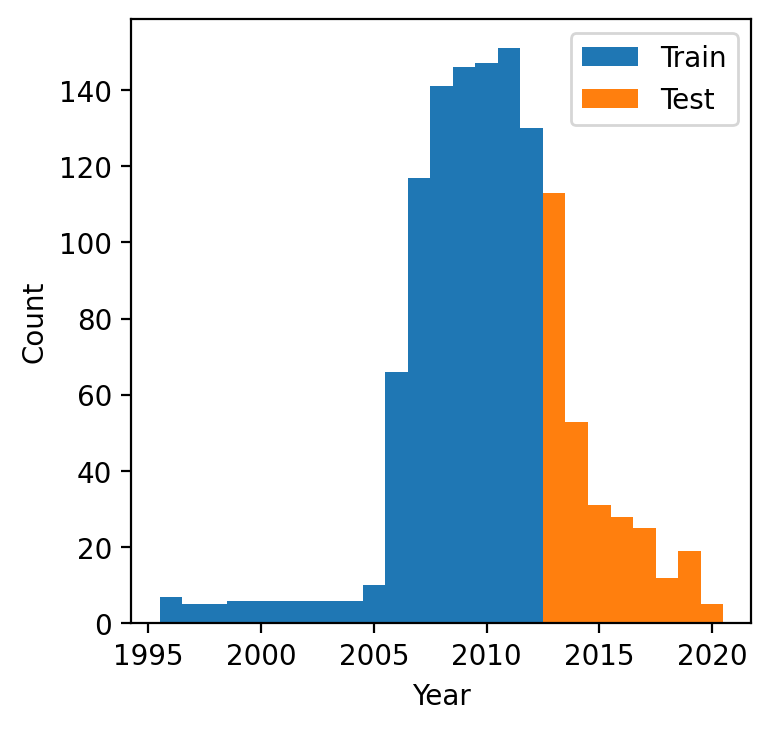}
         \caption{Training-testing split and number of observed MAGT values}
     \end{subfigure}
     \hfill
        \caption{Histograms of the number of observed ALT and MAGT values for a given year. The orange bars indicate the years that were used for the test.}
        \label{fig:tts}
\end{figure}

In total, we used 2 729 (ALT) and 961 (MAGT) data points, which were collected in 1990 -- 2010, from the areas of 364 000 km$^2$ and 322 000 km$^2$, respectively. These pixels on the map contain or are neighbors to the CALM sites and/or TSP boreholes.
To estimate the performance of our model and its ability to predict future values, we divided our dataset into training and test subsets. The training subsets preceded 2013, while the years for the test were 2013 -- 2020. 
Figure \ref{fig:tts} demonstrates an example of the training-testing split that we used.
Furthermore, we used the K-fold algorithm to divide the training set into five folds of 80\% randomly selected data points.

Tables ~\ref{tab:perm_results_alt} and ~\ref{tab:perm_results_tsoil} show the prediction errors for different models that were constructed using four sets of climatic factors with or without the outputs of Kudryavtsev models. 
We focus our evaluation of model performance on two metrics: the root mean square error (RMSE) ~\cite{Armstrong1992} and the coefficient of determination ($R^2$) ~\cite{fisher1921probable}.

The RMSE for Kudryavtsev model alone without using climate data was 32.39 $\pm$ 0.17 cm for the ALT and 1.26 $\pm$ 0.02 for the MAGT. This RMSE is a significant improvement over the existing approaches. When we accounted for all data in the \textbf{data} section, the test RMSE improved to 25.53 $\pm$ 0.40 cm for the ALT and 1.08 $\pm$ 0.02 for the MAGT.


\begin{table}
\begin{tabular}{llllll} 
\toprule
\hline
 &   & RMSE &  &  R$^2$ ($\times100$) & \\
Model & Factors  & Train &  Test  &   Train  &     Test \\ \hline
\midrule
CatBoost & All data &  0.72 $\pm$ .11 &  \textbf{1.08} $\pm$ .02 &    89 $\pm$ 3 &   \textbf{53} $\pm$ 1 \\
             & Kudr. (ALT + MAGT) &  1.04 $\pm$ .02 &  1.26 $\pm$ .02 &    77 $\pm$ 2 &   37 $\pm$ 2 \\
             & Kudr. (ALT) &  1.27 $\pm$ .03 &  1.32 $\pm$ .01 &    67 $\pm$ 2 &   30 $\pm$ 1 \\
             & Kudr. (MAGT) &  1.06 $\pm$ .08 &  1.22 $\pm$ .04 &    76 $\pm$ 3 &   41 $\pm$ 4 \\
             & Only climate data &  0.64 $\pm$ .06 &  1.14 $\pm$ .02 &    91 $\pm$ 2 &   48 $\pm$ 2 \\
              \hline
Elastic Net & All data &  1.24 $\pm$ .02 &  1.27 $\pm$ .03 &    68 $\pm$ 1 &   35 $\pm$ 3 \\
             & Kudr. (ALT + MAGT) &  1.50 $\pm$ .02 &  1.25 $\pm$ .01 &    53 $\pm$ 1 &   38 $\pm$ 1 \\
             & Kudr. (ALT) &  1.80 $\pm$ .04 &  1.50 $\pm$ .02 &    32 $\pm$ 1 &   10 $\pm$ 3 \\
             & Kudr. (MAGT) &  1.61 $\pm$ .02 &  1.40 $\pm$ .01 &    46 $\pm$ 1 &   21 $\pm$ 1 \\
             & Only climate data &  1.33 $\pm$ .02 &  1.55 $\pm$ .04 &    63 $\pm$ 1 &    4 $\pm$ 6 \\
              \hline
LinearRegression & All data &  1.18 $\pm$ .02 &  1.41 $\pm$ .07 &    71 $\pm$ 1 &   20 $\pm$ 8 \\
             & Kudr. (ALT + MAGT) &  1.42 $\pm$ .02 &  1.18 $\pm$ .02 &    58 $\pm$ 2 &   42 $\pm$ 2 \\
             & Kudr. (ALT) &  1.55 $\pm$ .03 &  1.19 $\pm$ .01 &    50 $\pm$ 2 &   44 $\pm$ 1 \\
             & Kudr. (MAGT) &  1.52 $\pm$ .02 &  1.17 $\pm$ .01 &    52 $\pm$ 1 &   45 $\pm$ 1 \\
             & Only climate data &  1.29 $\pm$ .02 &  1.64 $\pm$ .05 &    65 $\pm$ 1 &   -7 $\pm$ 7 \\
              \hline
NeuralNetwork& All data &  1.06 $\pm$ .07 &  1.56 $\pm$ .13 &    77 $\pm$ 3 &   2 $\pm$ 16 \\
             & Kudr. (ALT + MAGT) &  1.41 $\pm$ .03 &  1.39 $\pm$ .05 &    59 $\pm$ 2 &   23 $\pm$ 5 \\
             & Kudr. (ALT) &  1.50 $\pm$ .09 &  1.34 $\pm$ .01 &    53 $\pm$ 4 &   28 $\pm$ 1 \\
             & Kudr. (MAGT) &  1.45 $\pm$ .05 &  1.35 $\pm$ .11 &    57 $\pm$ 2 &  27 $\pm$ 11 \\
             & Only climate data &  1.05 $\pm$ .10 &  1.40 $\pm$ .23 &    77 $\pm$ 4 &  19 $\pm$ 26 \\
              \hline
RandomForest & All data &  0.38 $\pm$ .01 &  1.16 $\pm$ .03 &    \textbf{97} $\pm$ 1 &   47 $\pm$ 2 \\
             & Kudr. (ALT + MAGT) &  0.45 $\pm$ .02 &  1.24 $\pm$ .04 &    96 $\pm$ 1 &   38 $\pm$ 4 \\
             & Kudr. (ALT) &  0.51 $\pm$ .02 &  1.41 $\pm$ .04 &    94 $\pm$ 1 &   21 $\pm$ 5 \\
             & Kudr. (MAGT) &  0.48 $\pm$ .01 &  1.27 $\pm$ .04 &    95 $\pm$ 1 &   35 $\pm$ 4 \\
             & Only climate data &  \textbf{0.37} $\pm$ .01 &  1.28 $\pm$ .08 &    97 $\pm$ 1 &   35 $\pm$ 8 \\
      \hline
\bottomrule
\end{tabular}

    \caption{Mean annual ground temperatures (MAGT) prediction errors on the proposed training-testing splits. For each configuration, we trained the model with 80\% of the training data and tested it on the test data. The values show the average of five estimations for each model configuration. Here, Kudr. denotes the Kudryavtsev model results. The best result is shown in bold.}
    \label{tab:perm_results_alt}
\end{table}

\begin{table}
\begin{tabular}{llllll} 
\toprule
\hline
 &   & RMSE &  &  R$^2$ ($\times100$)& \\
Model & Factors  & Train &  Test  &   Train  &     Test \\ \hline
\midrule
CatBoost & All data &    \textbf{5.65} $\pm$ .90 &   \textbf{25.53} $\pm$ .40 &    \textbf{98} $\pm$ 1 &   \textbf{62} $\pm$ 1 \\
             & Kudr. (ALT + MAGT) &   27.32 $\pm$ .55 &   32.39 $\pm$ .17 &    61 $\pm$ 2 &   38 $\pm$ 1 \\
             & Kudr. (ALT) &  25.71 $\pm$ 1.69 &   36.71 $\pm$ .13 &    65 $\pm$ 5 &   21 $\pm$ 1 \\
             & Kudr. (MAGT) &  27.41 $\pm$ 1.48 &   33.62 $\pm$ .54 &    60 $\pm$ 4 &   33 $\pm$ 2 \\
             & Only climate data &   9.96 $\pm$ 1.63 &   28.57 $\pm$ .47 &    95 $\pm$ 2 &   52 $\pm$ 2 \\
             \hline
Elastic Net & All data &   29.80 $\pm$ .31 &   37.79 $\pm$ .18 &    53 $\pm$ 1 &   16 $\pm$ 1 \\
             & Kudr. (ALT + MAGT) &   38.93 $\pm$ .39 &   40.17 $\pm$ .18 &    20 $\pm$ 1 &    5 $\pm$ 1 \\
             & Kudr. (ALT) &   39.79 $\pm$ .35 &   40.80 $\pm$ .39 &    17 $\pm$ 1 &    2 $\pm$ 2 \\
             & Kudr. (MAGT) &   39.48 $\pm$ .40 &   39.96 $\pm$ .18 &    18 $\pm$ 1 &    6 $\pm$ 1 \\
             & Only climate data &   31.14 $\pm$ .28 &   38.48 $\pm$ .27 &    49 $\pm$ 1 &   13 $\pm$ 1 \\
             \hline
LinearRegression & All data &   28.48 $\pm$ .23 &   36.02 $\pm$ .23 &    57 $\pm$ 1 &   24 $\pm$ 1 \\
             & Kudr. (ALT + MAGT) &   36.43 $\pm$ .27 &   39.76 $\pm$ .38 &    30 $\pm$ 1 &    7 $\pm$ 2 \\
             & Kudr. (ALT) &   39.13 $\pm$ .38 &   40.76 $\pm$ .15 &    20 $\pm$ 1 &    2 $\pm$ 1 \\
             & Kudr. (MAGT) &   38.23 $\pm$ .36 &   39.62 $\pm$ .33 &    23 $\pm$ 1 &    7 $\pm$ 2 \\
             & Only climate data &   31.07 $\pm$ .28 &   38.93 $\pm$ .28 &    49 $\pm$ 1 &   11 $\pm$ 1 \\
             \hline
NeuralNetwork & All data &  19.21 $\pm$ 1.30 &   29.53 $\pm$ .60 &    81 $\pm$ 3 &   49 $\pm$ 2 \\
             & Kudr. (ALT + MAGT) &   33.50 $\pm$ .62 &  39.31 $\pm$ 1.88 &    41 $\pm$ 3 &    9 $\pm$ 9 \\
             & Kudr. (ALT) &   38.65 $\pm$ .48 &   38.35 $\pm$ .31 &    21 $\pm$ 1 &   13 $\pm$ 1 \\
             & Kudr. (MAGT) &   31.13 $\pm$ .94 &   36.73 $\pm$ .28 &    49 $\pm$ 3 &   21 $\pm$ 1 \\
             & Only climate data &  17.81 $\pm$ 2.89 &  32.30 $\pm$ 1.14 &    83 $\pm$ 5 &   38 $\pm$ 4 \\
             \hline
RandomForest & All data &    7.36 $\pm$ .12 &   29.57 $\pm$ .32 &    97 $\pm$ 1 &   48 $\pm$ 1 \\
             & Kudr. (ALT + MAGT) &   10.43 $\pm$ .16 &   34.53 $\pm$ .26 &    94 $\pm$ 1 &   30 $\pm$ 1 \\
             & Kudr. (ALT) &   12.89 $\pm$ .12 &   37.05 $\pm$ .16 &    91 $\pm$ 1 &   19 $\pm$ 1 \\
             & Kudr. (MAGT) &   11.58 $\pm$ .28 &   35.03 $\pm$ .44 &    93 $\pm$ 1 &   28 $\pm$ 2 \\
             & Only climate data &    7.62 $\pm$ .12 &   31.70 $\pm$ .28 &    97 $\pm$ 1 &   41 $\pm$ 1 \\
      \hline
\bottomrule
\end{tabular}

    \caption{Active layer thickness (ALT) prediction errors on the training-testing splits. For each configuration, we trained the model with 80\% of the training data and tested it on the test data. The values show the average of five estimations for each model configuration. Here, Kudr. denotes the Kudryavtsev model results. The best result is shown in bold.}
    \label{tab:perm_results_tsoil}
\end{table}

\begin{figure}
     \centering
     \begin{subfigure}[b]{7.7cm}
         \includegraphics[width=7.7cm]{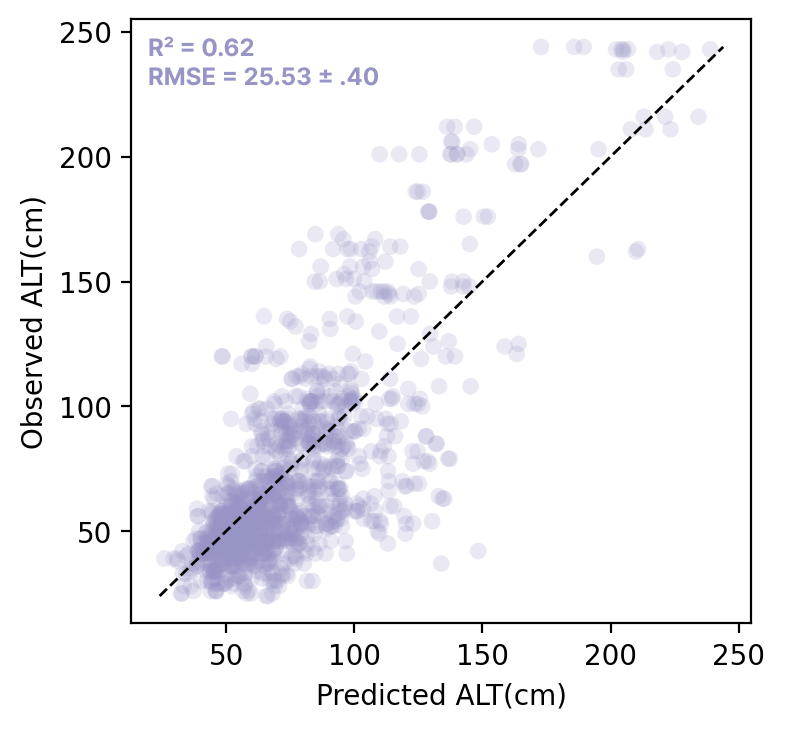}
         \caption{Predicted \textit{vs} actual ALT.}
     \end{subfigure}
     \hfill
     \begin{subfigure}[b]{7.7cm}
         \includegraphics[width=7.5cm]{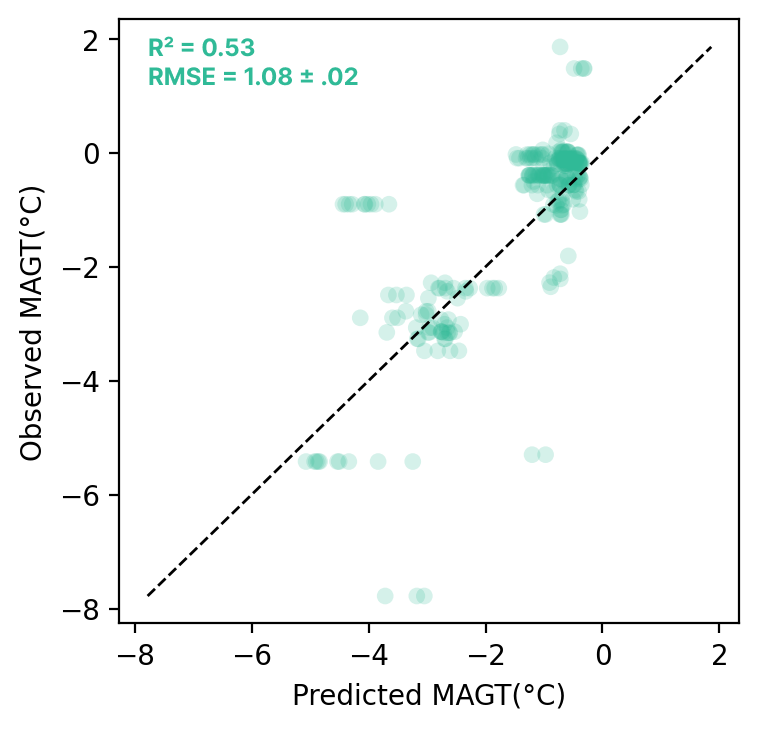}
         \caption{Predicted \textit{vs} actual MAGT}
     \end{subfigure}
     \hfill
        \caption{Scatter plots of the predicted \textit{vs} actual values for the active layer thickness and mean annual ground temperature. }
        \label{fig:alt_pred_vs_act}
\end{figure}

Figure \ref{fig:alt_pred_vs_act} shows the predicted values by the CatBoost model, which was trained with all factors included \textit{versus} the actual active layer thickness computed using all climatic data in the \textbf{Data} section. Although our model demonstrates excellent predictive power, it slightly underestimates the ALT.

\begin{figure}[h!]
    \centering
    \includegraphics[width=0.90\textwidth]{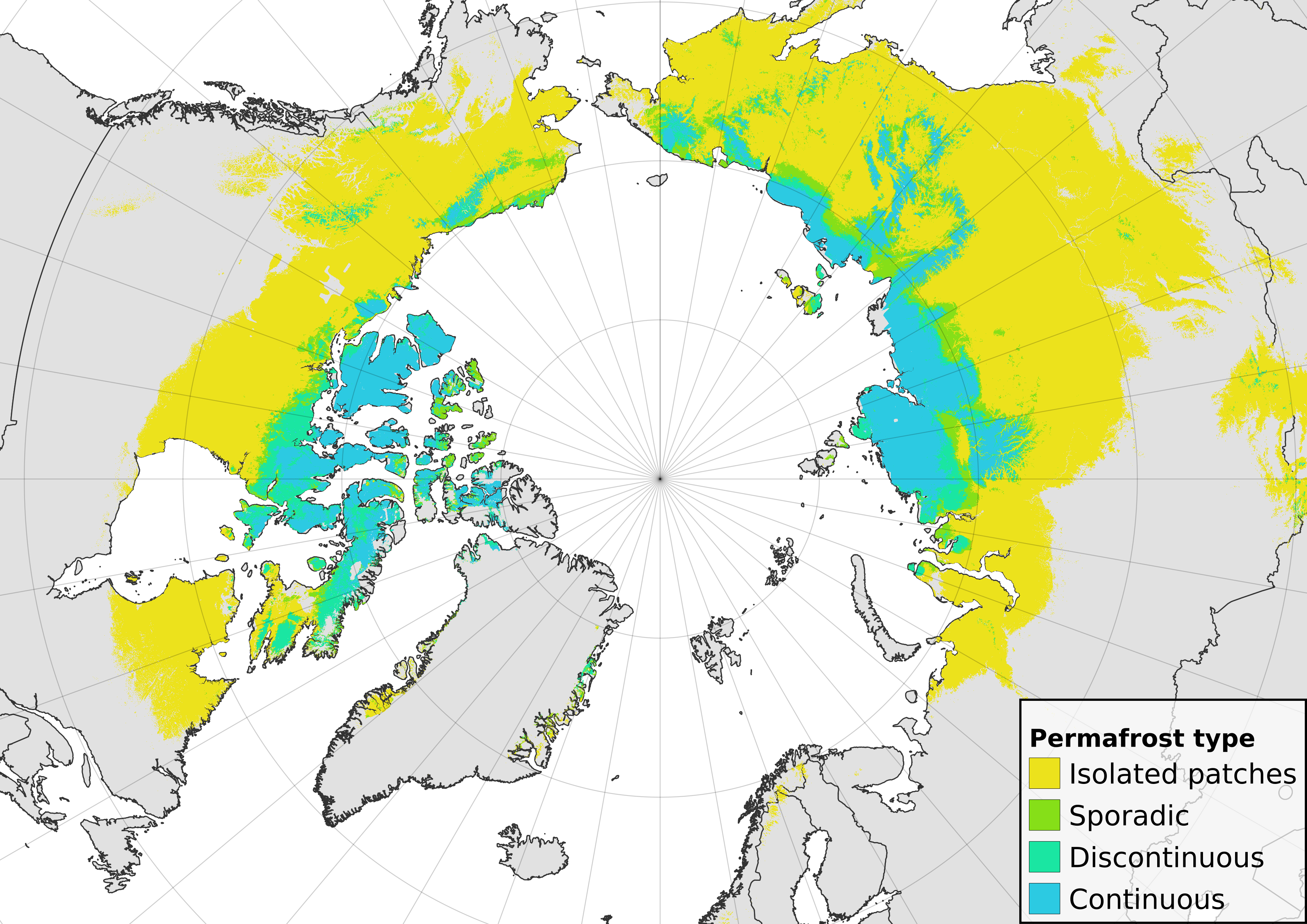}
    \caption{Predicted state of permafrost for 2050 under the CMIP 6 SSP245 scenario.}
    \label{fig:predicted_2050}
\end{figure}

We applied our model to predict the extent of permafrost degradation under the optimistic CMIP 6 SSP245 scenario (updated RCP 4.5 \cite{riahi2017shared}). The prediction for the permafrost degradation by 2050 is demonstrated in Figure \ref{fig:predicted_2050}. 

\begin{figure}
     \centering
     \begin{subfigure}[b]{0.49\textwidth}
         \centering
         \includegraphics[width=\textwidth]{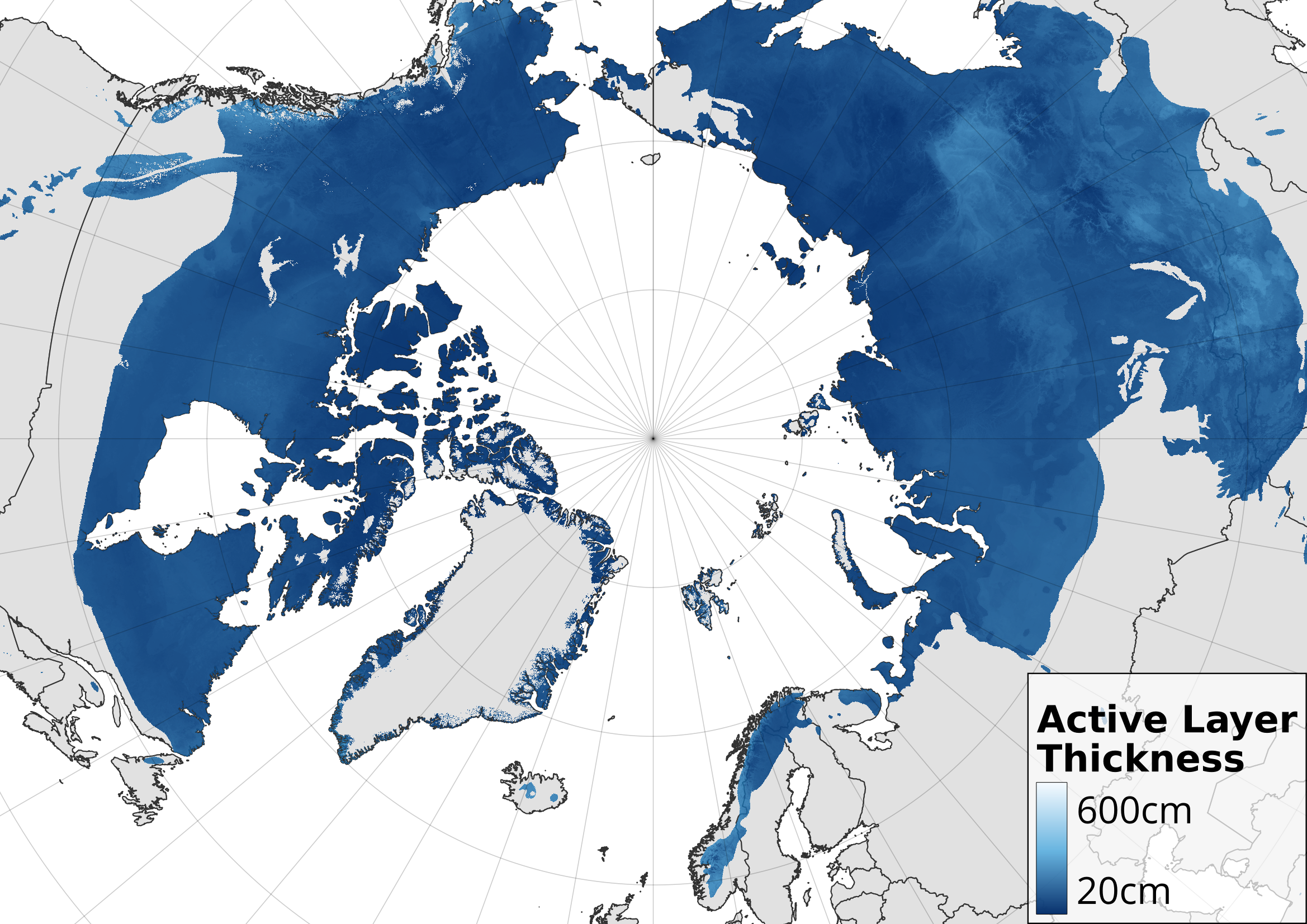}
         \caption{Active layer thickness, 2010}
     \end{subfigure}
     \hfill
     \begin{subfigure}[b]{0.49\textwidth}
         \centering
         \includegraphics[width=\textwidth]{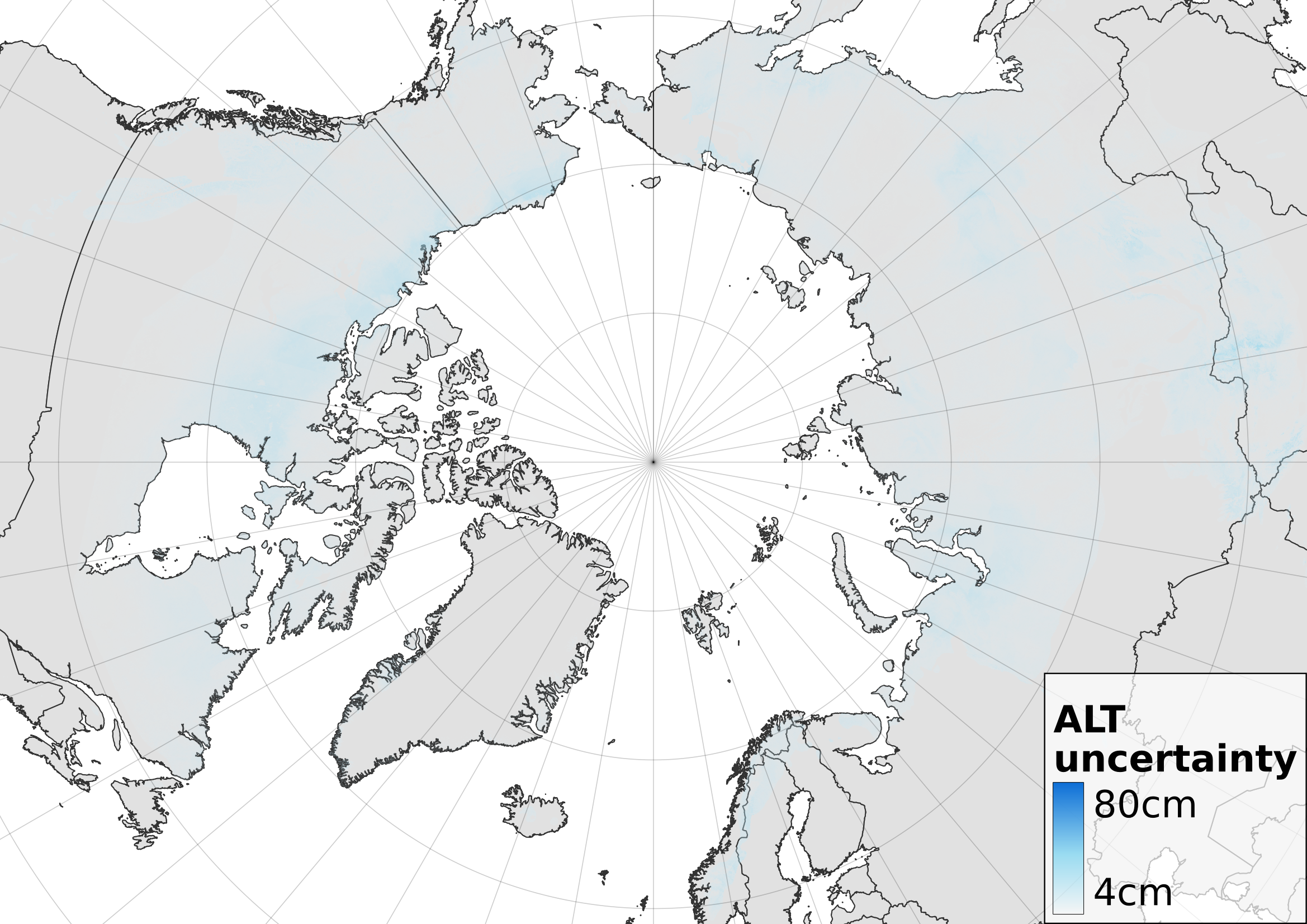}
         \caption{Active layer thickness uncertainty, 2010}
     \end{subfigure}
     \hfill
     \\
     \begin{subfigure}[b]{0.49\textwidth}
         \centering
         \includegraphics[width=\textwidth]{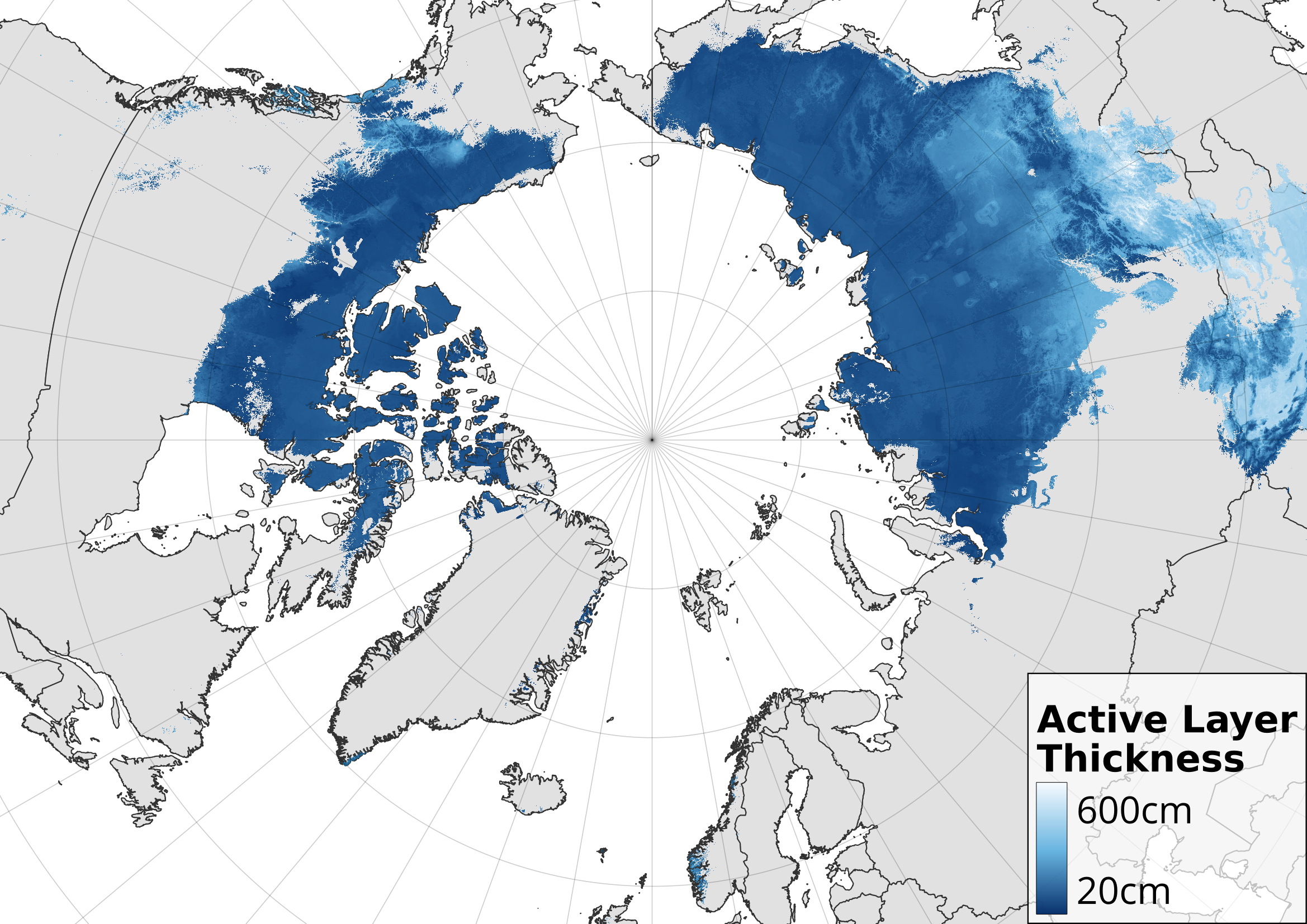}
         \caption{Active layer thickness, 2050}
     \end{subfigure}
     \hfill
     \begin{subfigure}[b]{0.49\textwidth}
         \centering
         \includegraphics[width=\textwidth]{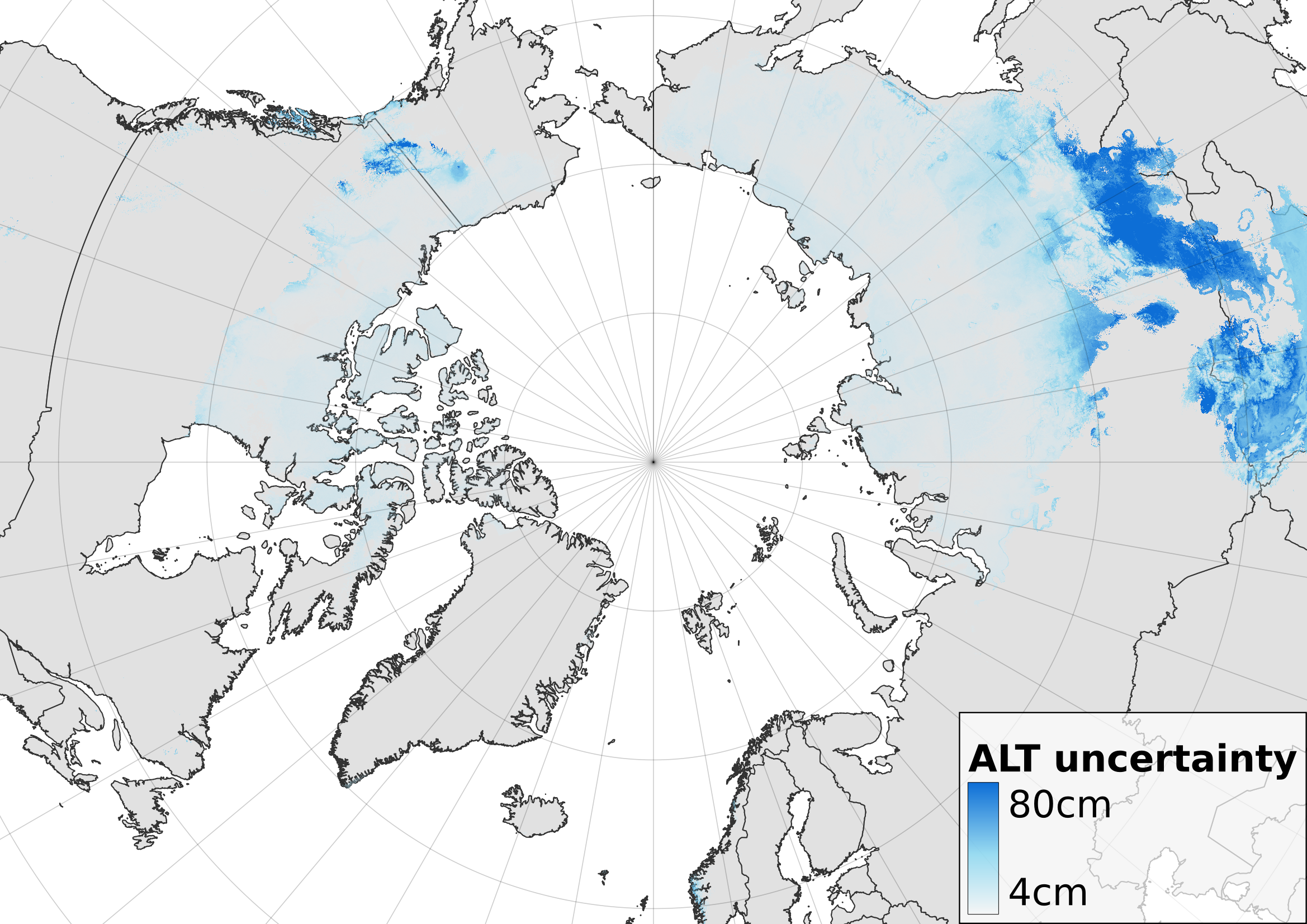}
         \caption{Active layer thickness uncertainty, 2050}
     \end{subfigure}
     \hfill
     \\
     \begin{subfigure}[b]{0.49\textwidth}
         \centering
         \includegraphics[width=\textwidth]{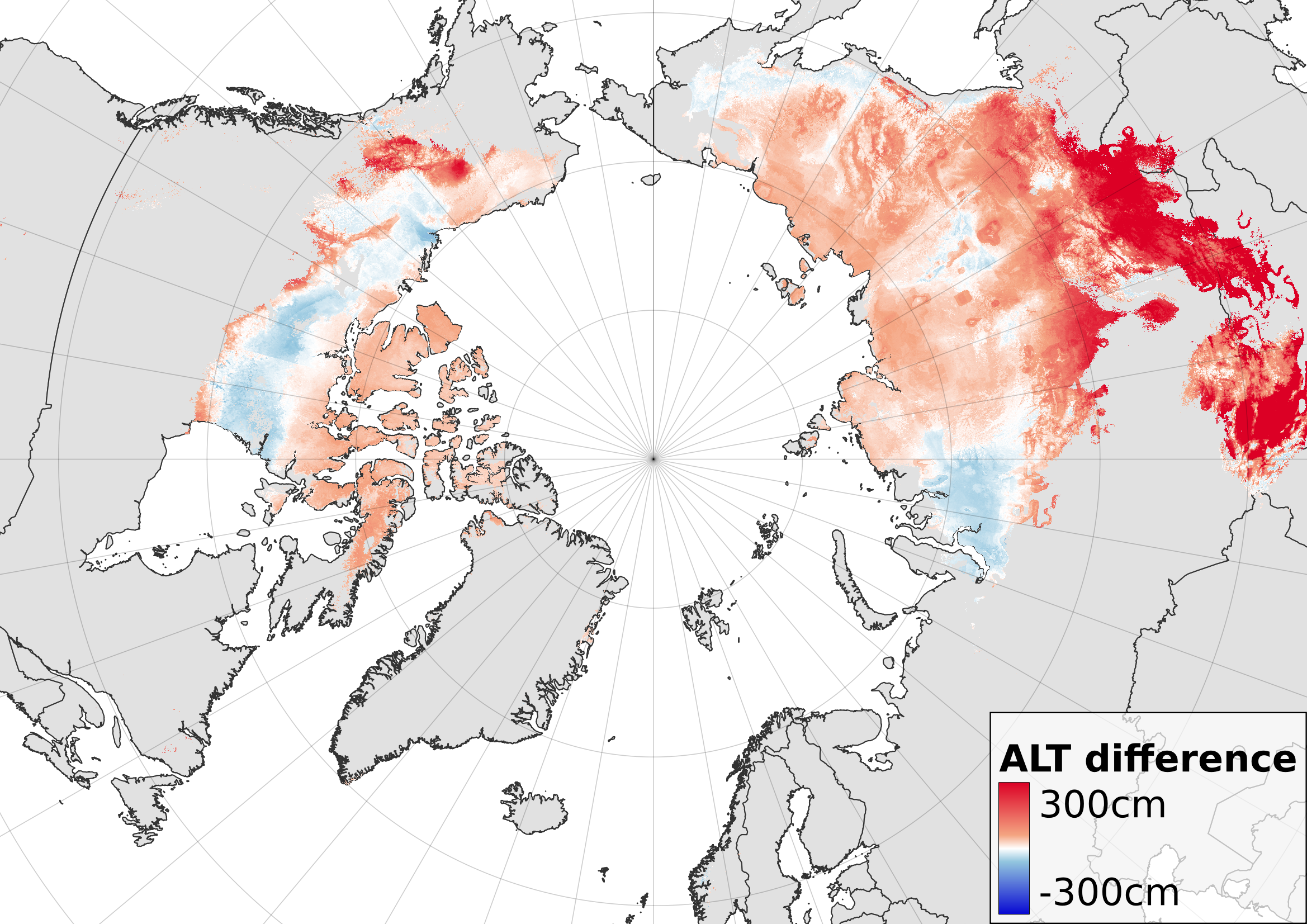}
         \caption{Active layer thickness difference between 2050 and 2010}
     \end{subfigure}
     \hfill
     \begin{subfigure}[b]{0.49\textwidth}
         \centering
         \includegraphics[width=\textwidth]{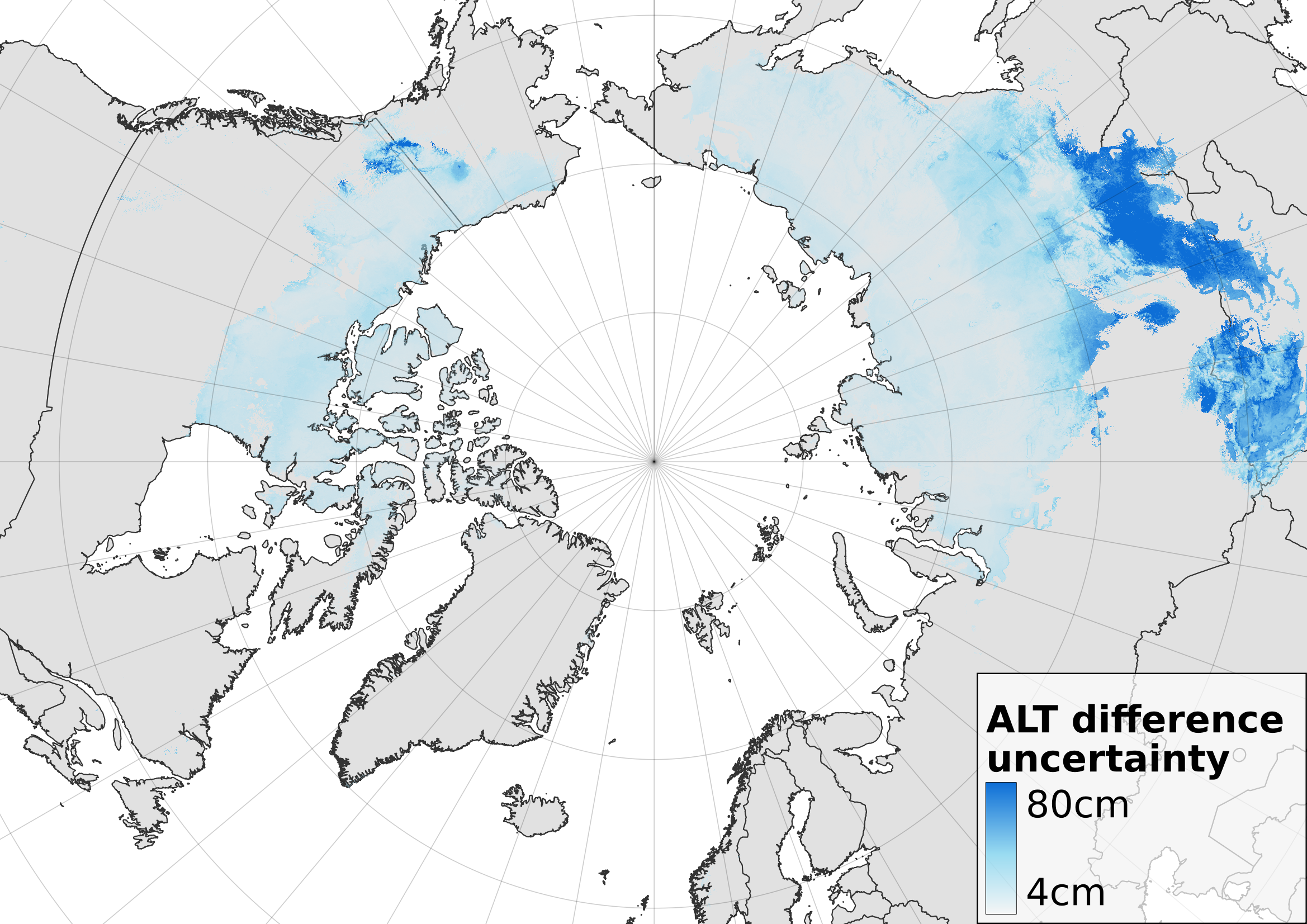}
         \caption{Uncertainty for the difference in active layer thickness between 2050 and 2010}
     \end{subfigure}
     \hfill
        \caption{Comparison of the active layer thickness in 2010 and 2050 under the CMIP6 SSP245 scenario.}
        \label{fig:alt_pred2050}
\end{figure}

\begin{figure}
     \centering
     \begin{subfigure}[b]{0.49\textwidth}
         \centering
         \includegraphics[width=\textwidth]{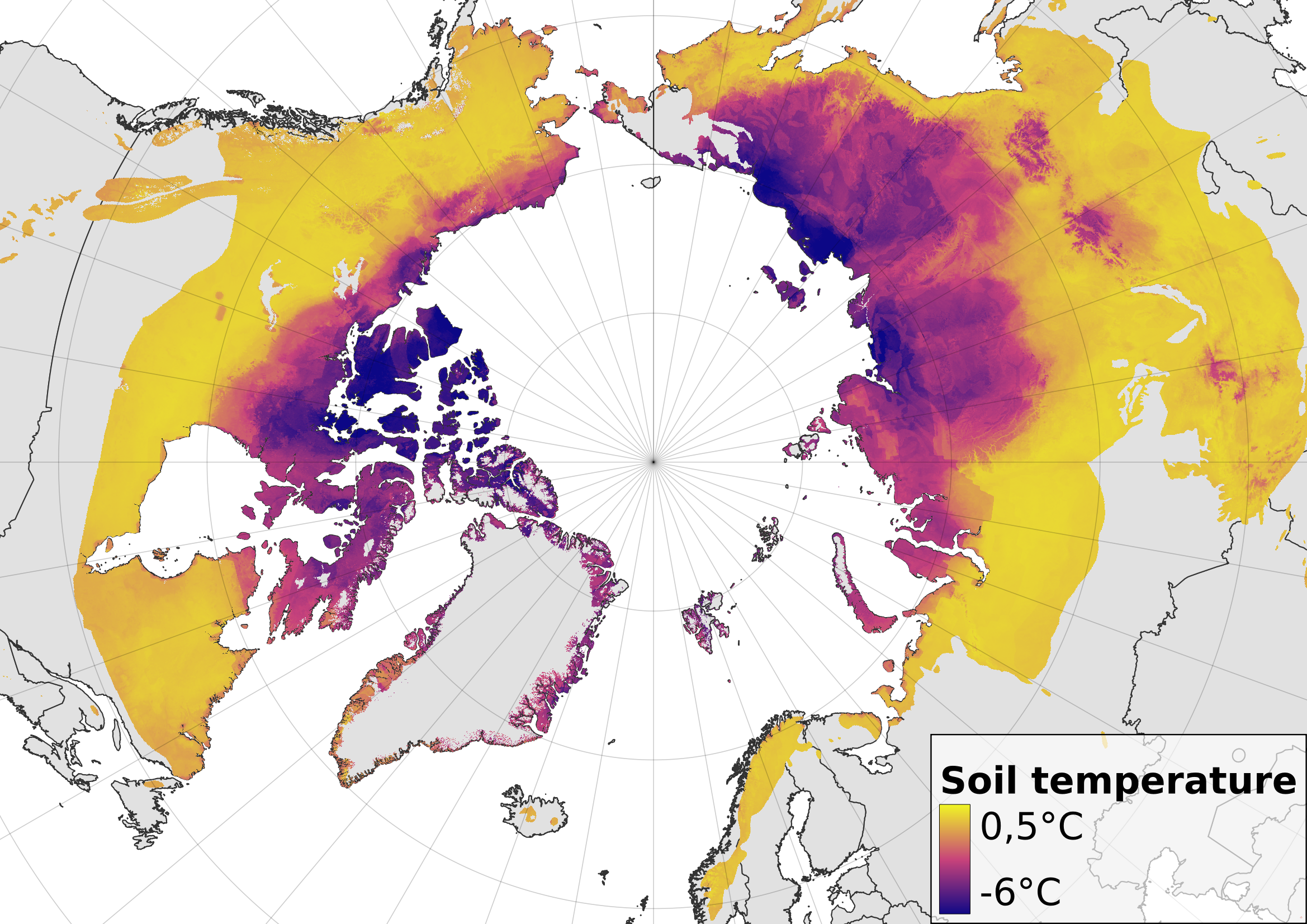}
         \caption{Temperature, 2010}
     \end{subfigure}
     \hfill
     \begin{subfigure}[b]{0.49\textwidth}
         \centering
         \includegraphics[width=\textwidth]{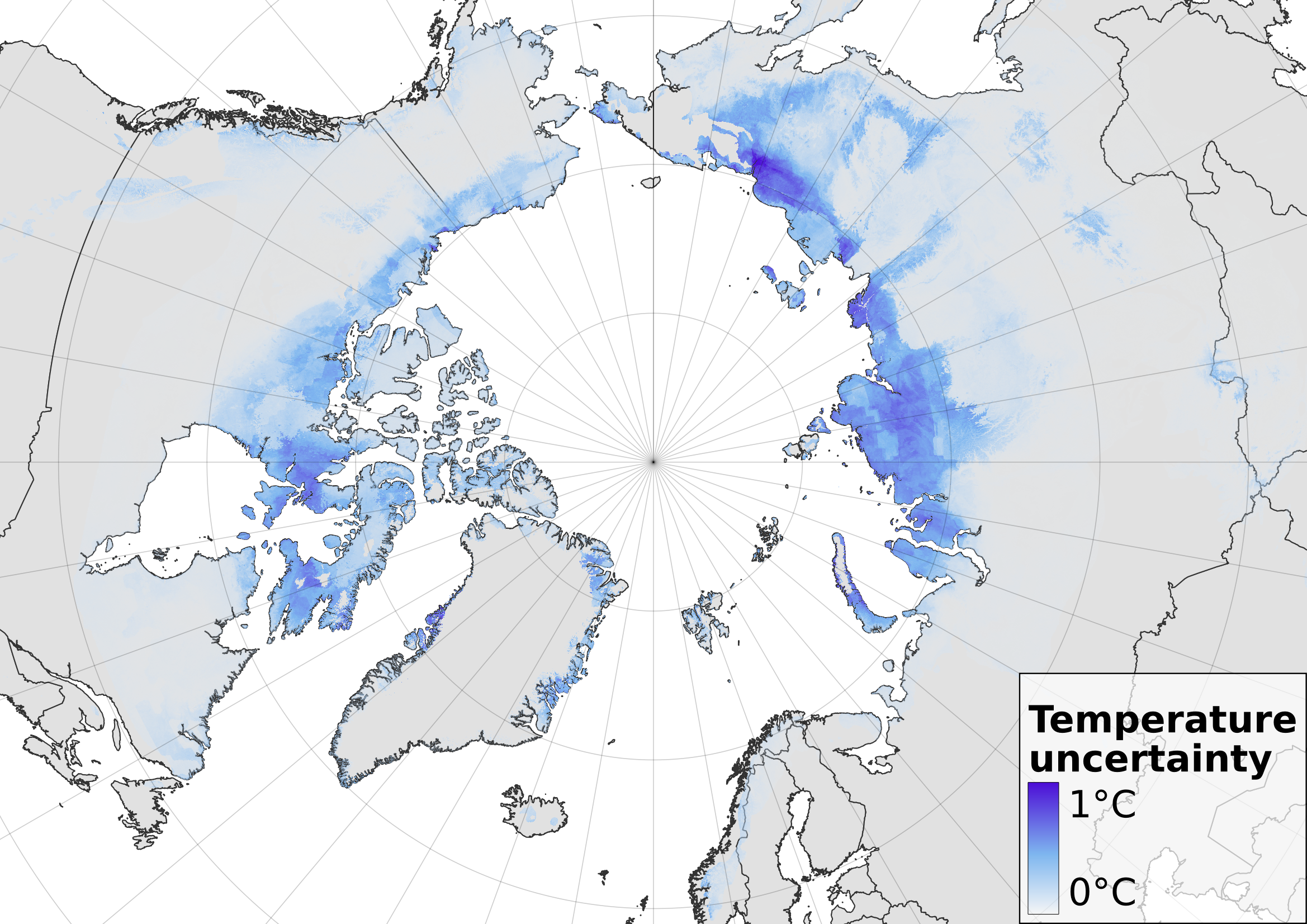}
         \caption{Temperature uncertainty, 2010}
     \end{subfigure}
     \hfill
     \\
     \begin{subfigure}[b]{0.49\textwidth}
         \centering
         \includegraphics[width=\textwidth]{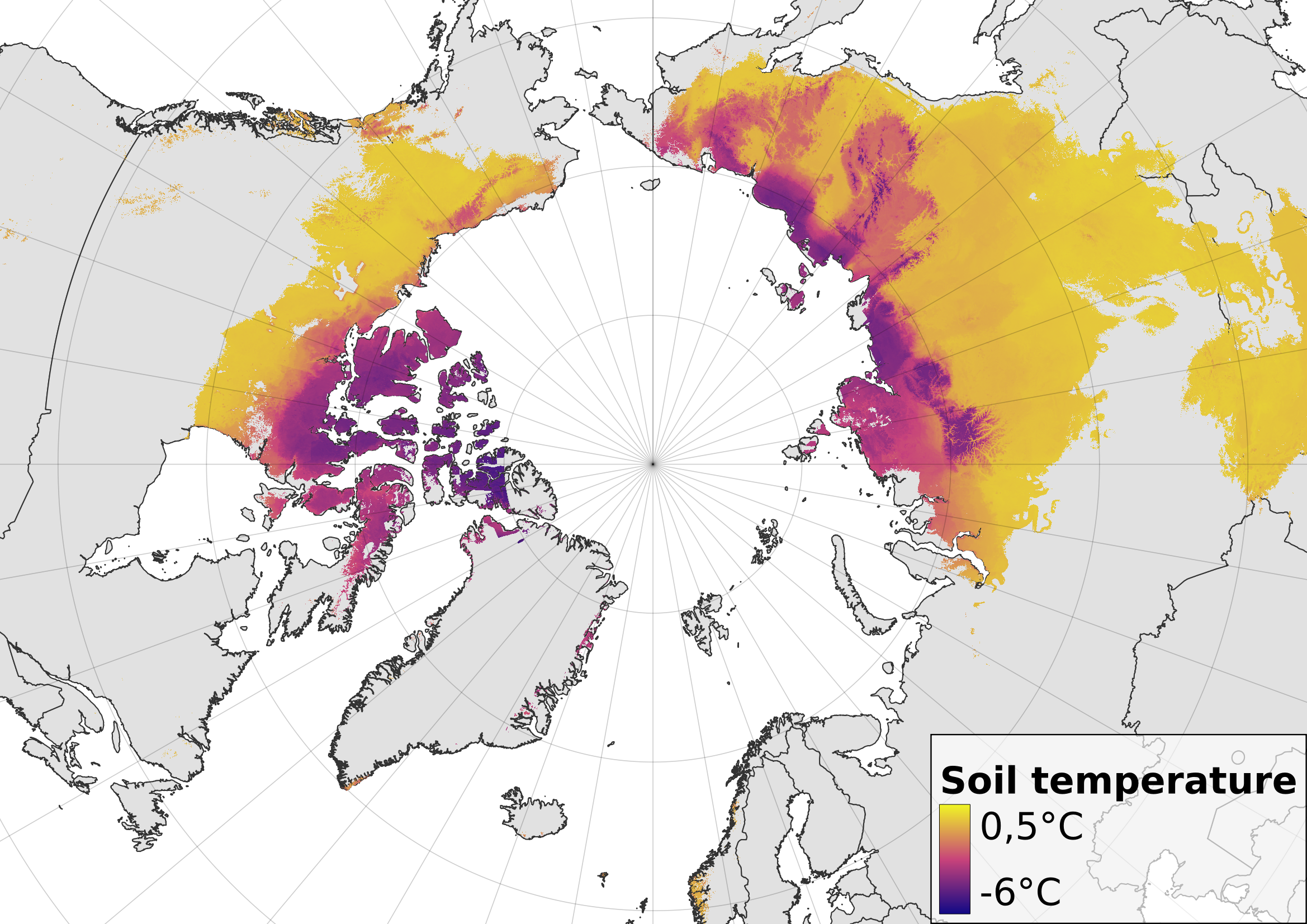}
         \caption{Temperature, 2050}
     \end{subfigure}
     \hfill
     \begin{subfigure}[b]{0.49\textwidth}
         \centering
         \includegraphics[width=\textwidth]{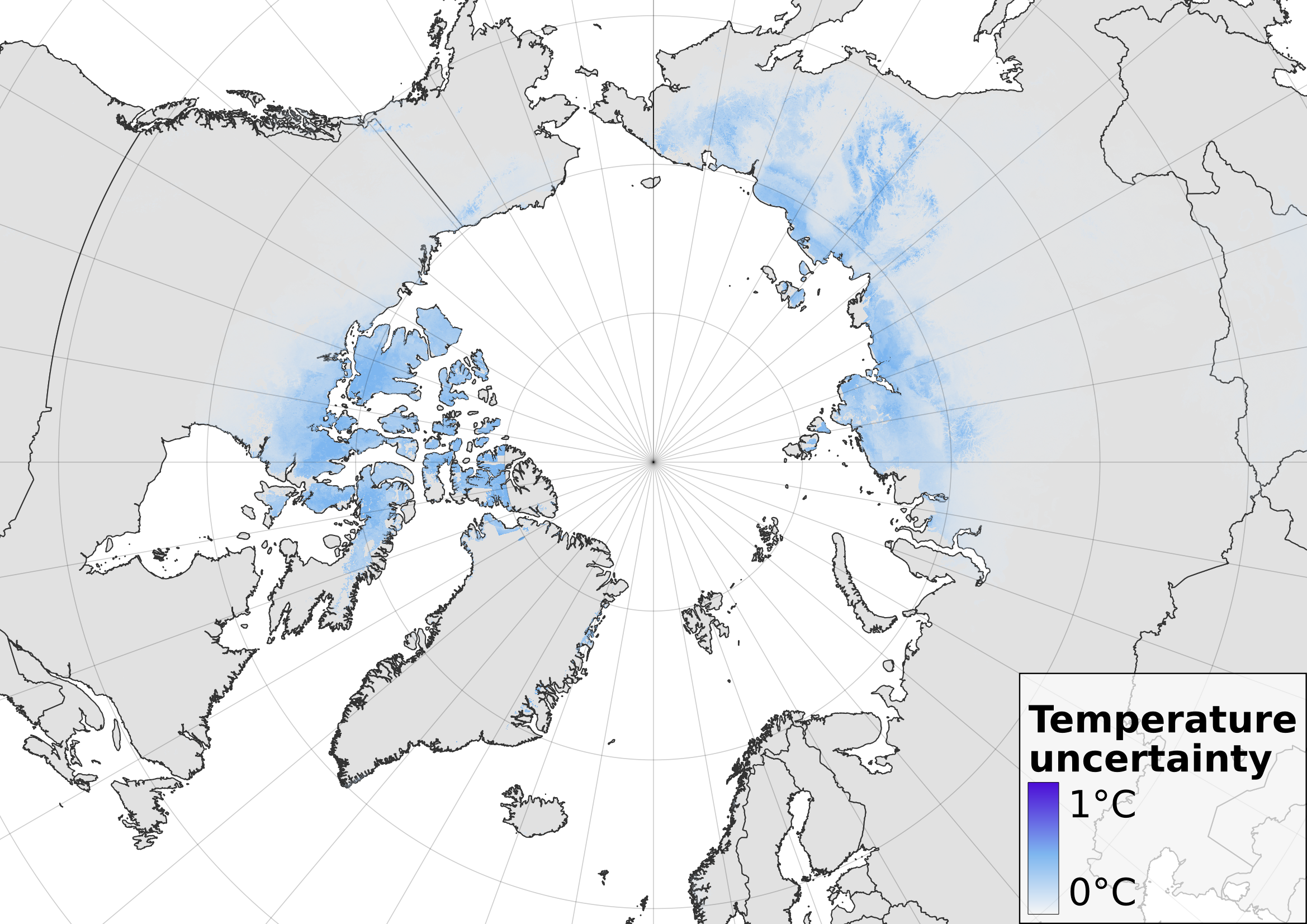}
         \caption{Temperature uncertainty, 2050}
     \end{subfigure}
     \hfill
     \\
     \begin{subfigure}[b]{0.49\textwidth}
         \centering
         \includegraphics[width=\textwidth]{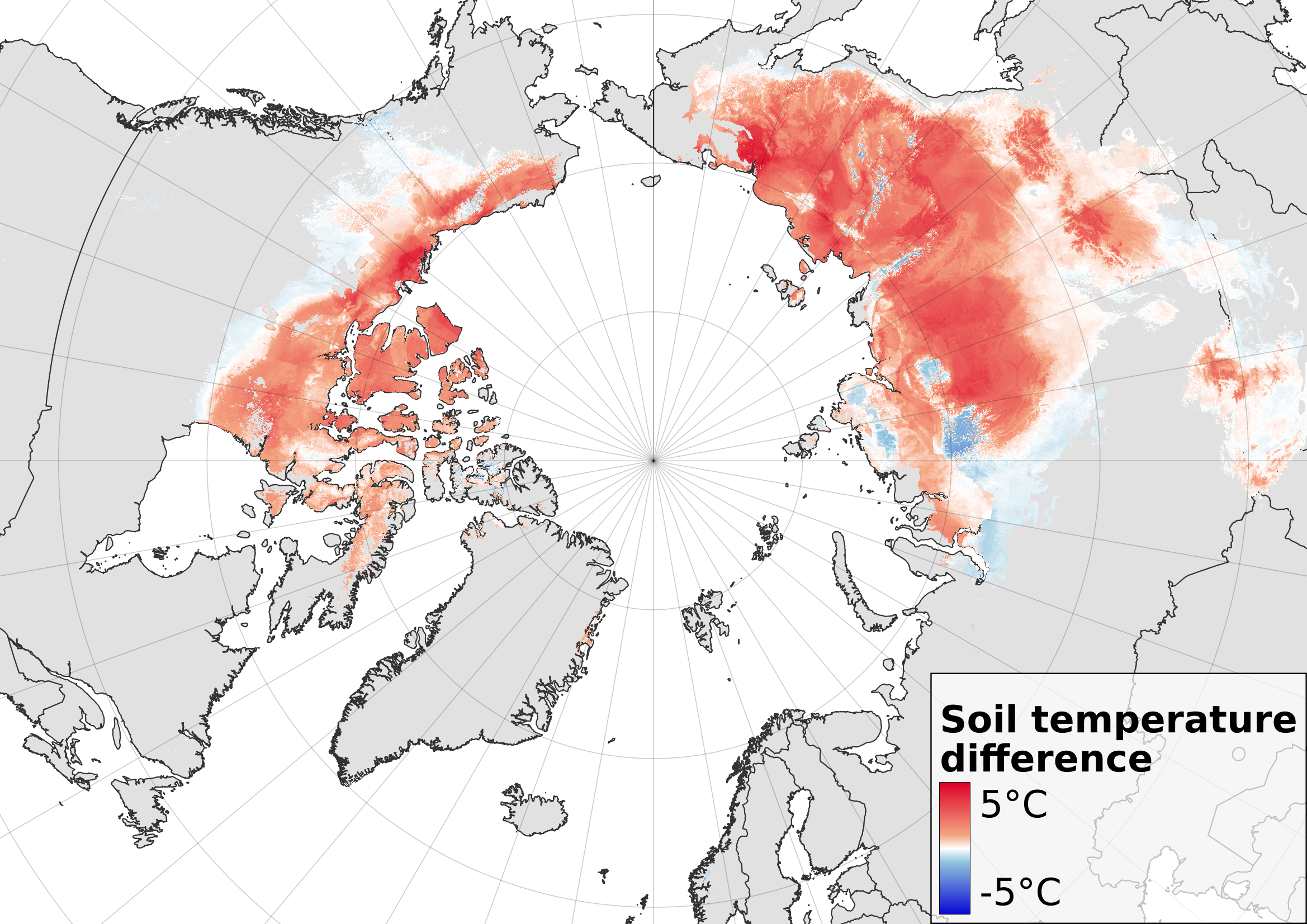}
         \caption{Temperature difference between 2050 and 2010}
     \end{subfigure}
     \hfill
     \begin{subfigure}[b]{0.49\textwidth}
         \centering
         \includegraphics[width=\textwidth]{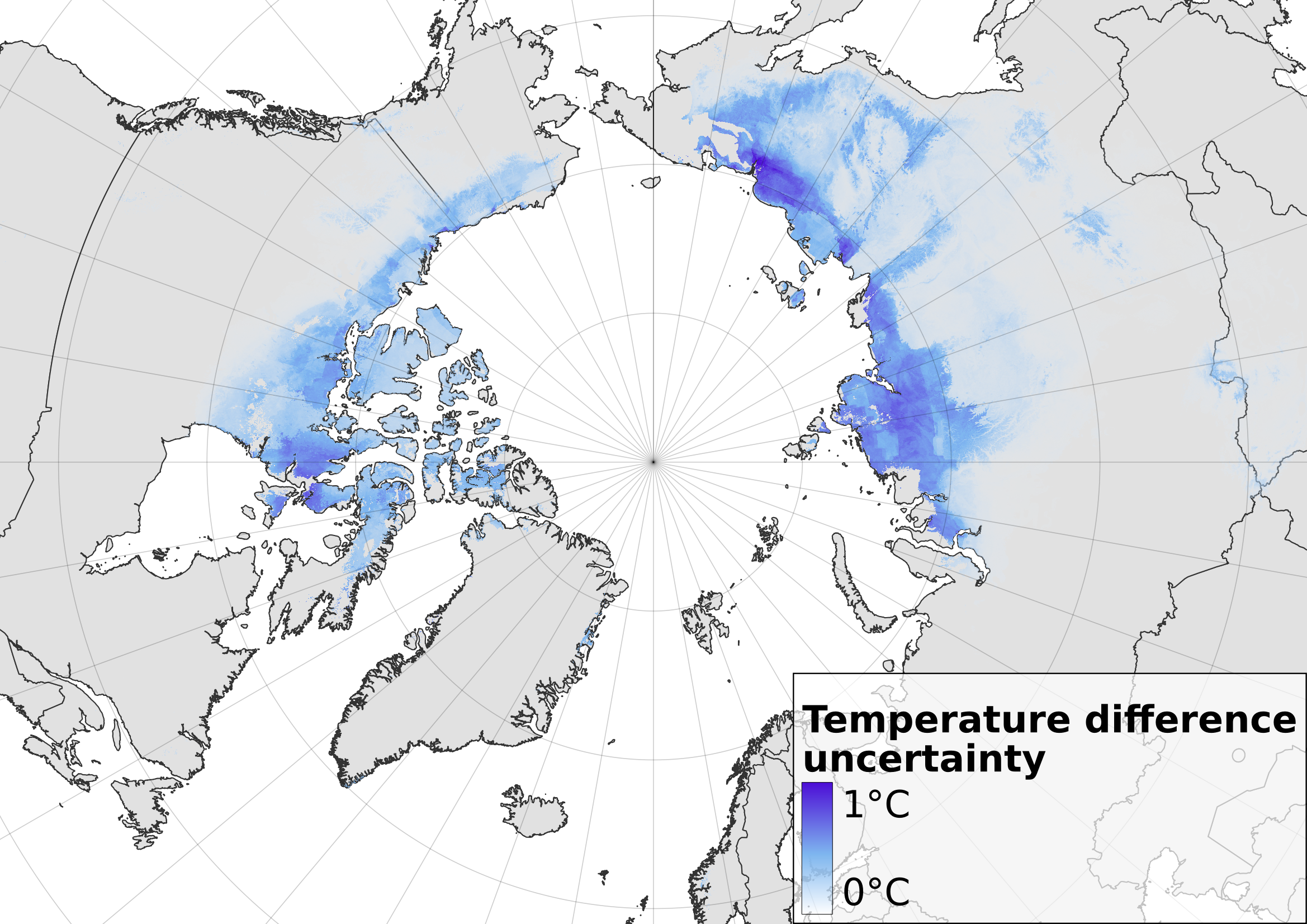}
         \caption{Uncertainty for the temperature difference between 2050 and 2010}
     \end{subfigure}
     \hfill
        \caption{ Comparison of the mean annual ground temperature at zero amplitude in 2010 and 2050 under the CMIP 6 SSP245 scenario.}
        \label{fig:TEMP_pred2050}
\end{figure}

Figures \ref{fig:alt_pred2050} and \ref{fig:TEMP_pred2050} demonstrate the predicted active layer thickness and temperature of the soil at the zero-amplitude level for 2050 (under the CMIP 6 SSP245 scenario) and as a comparison to 2010.

\section*{Discussion}

Significant progress in machine learning algorithms and methods for natural sciences cannot be ignored while studying climate change and permafrost degradation as a key part of it. While promising in predicting environmental changes, machine learning methods demonstrate mediocre quality over sparse, limited, and incomplete data. The latter is intrinsic for sparsely populated Arctic areas and permafrost melting there. 

The major challenge in data-driven permafrost modeling and thaw depth prediction comes from sparse and irregular data on the active layer thickness dynamics. Moreover, the locations of CALM sites do not cover the actual temperature range for all types of landscapes or soils. 


As such, classical physics-based models are still dominant in permafrost thaw and environmental modeling in the Arctic. In this paper, we are the first to demonstrate the power of machine learning algorithms properly augmented by physical models for permafrost analysis and long-term thaw prediction. In particular, we show that a coarse-grained physical model, such as a Kudryavtsev solution to the heat equation \cite{kudryavtsev1977fundamentals}, used as a regularization to a machine learning method can improve the performance of both machine learning and physics-based models by 20\% or more. 

We believe the latter opens the door to studying a wide range of environmental processes in high latitudes from the physics-informed machine learning viewpoint, as the latter may substantially improve existing algorithms and practices. Furthermore, it will provide stakeholders with a high-fidelity tool to mitigate the climate change impact and protect people and the planet.


\newpage


\section*{Supplementary materials}
Figs. S1-S2\\




\bibliography{scibib}
\bibliographystyle{Science}

\newpage
\section*{Supplementary materials}

\textbf{The PDF file includes:}
\begin{quote}
Figures S1 to S2\\
\end{quote}

\renewcommand{\figurename}{Supplementary Figure}
\setcounter{figure}{0}

\begin{figure}[h]
    \centering
    \includegraphics[width=16cm]{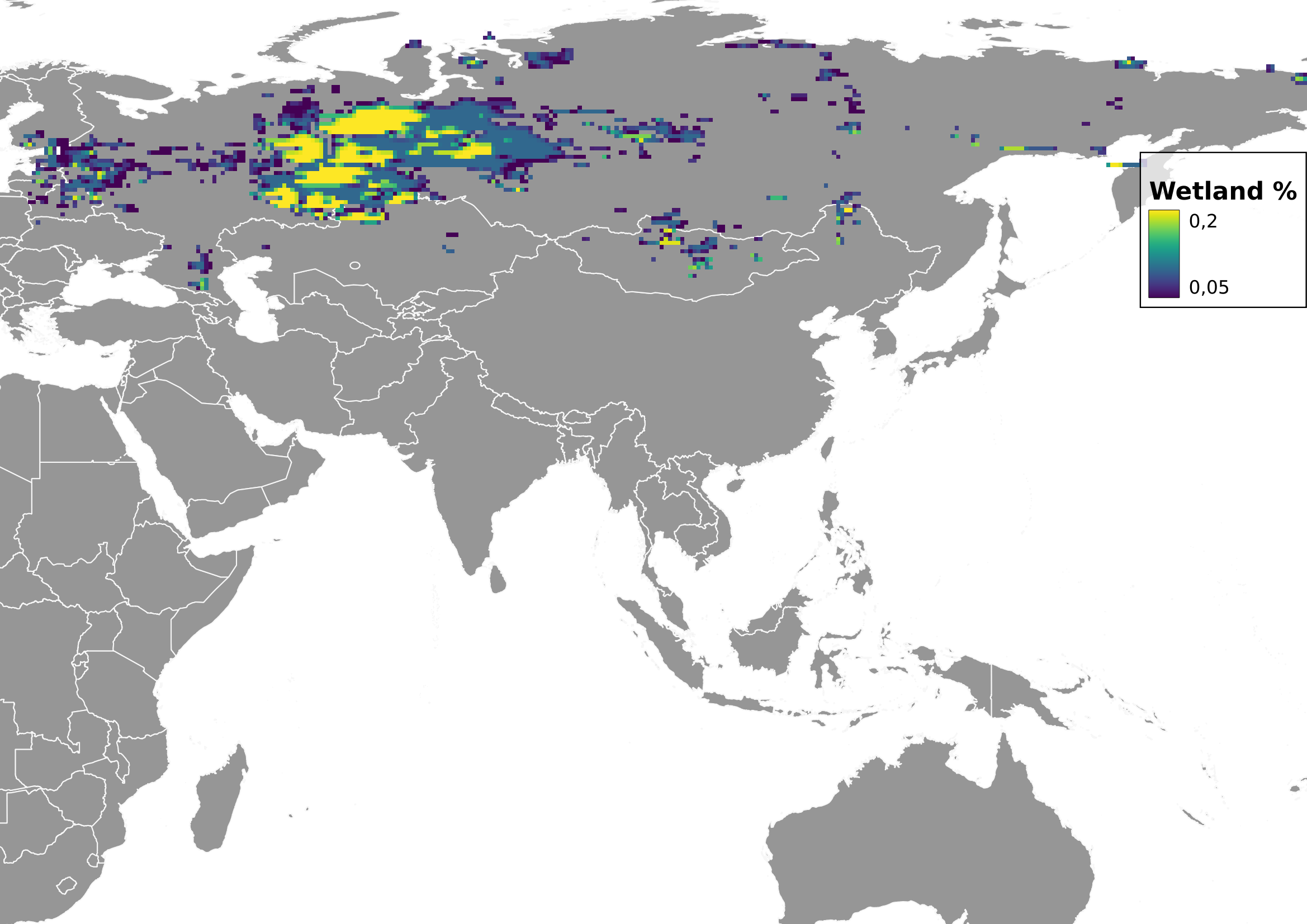}
    \caption{Proportion of wetlands in each pixel.}
    \label{fig:swamp}
\end{figure}

\begin{figure}[h]
    \centering
    \includegraphics[width=16cm]{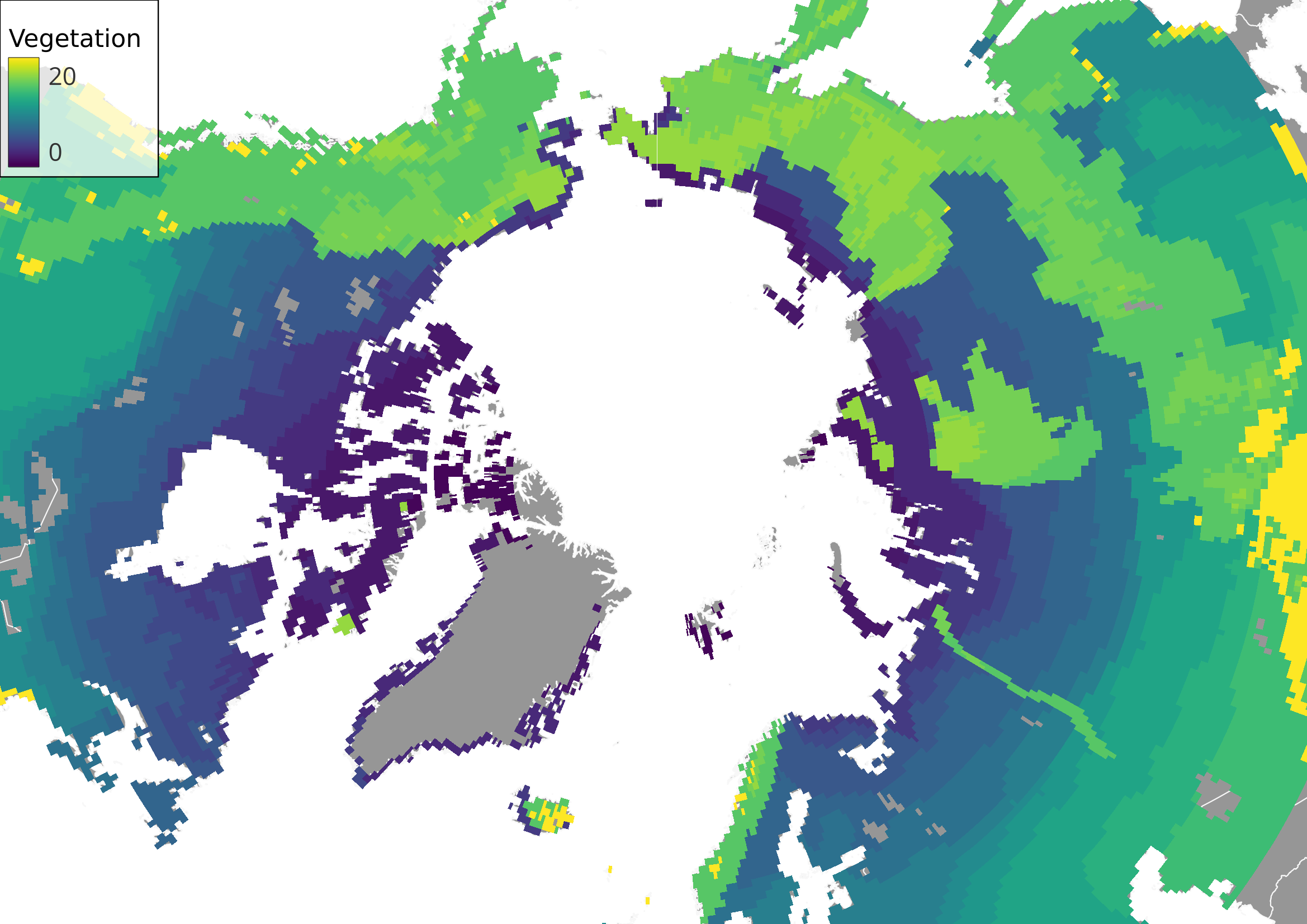}
    \caption{One-dimensional representation of the vegetation types.}
    \label{fig:veg}
\end{figure}

\end{document}